\documentclass[twocolumn,final,epj]{svjour}
\usepackage[utf8]{inputenc}
\usepackage[T1]{fontenc}
\usepackage{graphicx}
\usepackage{amssymb}
\usepackage{amsmath}
\usepackage{hyperref}
\usepackage{color}
\definecolor{blue-violet}{rgb}{0.54, 0.17, 0.89}
\definecolor{ao(english)}{rgb}{0.0, 0.5, 0.0}

\begin{document}

\title{Google matrix of Bitcoin network}

\author{
Leonardo Ermann\inst{1}, 
Klaus M. Frahm\inst{2} \and 
Dima L. Shepelyansky\inst{2}}

\institute{
Departamento de F\'\i sica Te\'orica, GIyA, Comisi\'on Nacional 
de Energ\'ia At\'omica, Buenos Aires, Argentina
\and
Laboratoire de Physique Th\'eorique du CNRS, IRSAMC, 
Universit\'e de Toulouse, CNRS, UPS, 31062 Toulouse, France
}


\date{\today}

\abstract{We construct and study the Google matrix of Bitcoin transactions during the time period
from the very beginning in 2009 till April 2013. The Bitcoin network has up to a few millions of bitcoin users
and we present its main characteristics including the PageRank and CheiRank probability distributions,
the spectrum of eigenvalues of Google matrix and related eigenvectors.
We find that the spectrum has an unusual circle-type structure which we attribute to 
existing hidden communities of nodes linked between their members.
We show that  the Gini coefficient of the transactions for the whole period
is close to unity showing that the main part of wealth of the network is 
captured by a small fraction of users.  
}


\maketitle

\section{Introduction}

The bitcoin crypto currency was introduced by Satoshi 
Nakamoto in 2009 \cite{nakamoto} and became at present an important source of 
direct financial exchange between private users \cite{wikibitcoin}.
At present this new cryptographic manner of financial exchange
attracts a significant interest of society, computer scientists, economists and politicians
(see e.g. \cite{bitcoinscience,biryukov,fedlet,samuk,cryptocurrency}). 
The amazing feature of bitcoin transactions is that all of them are
open to public at \cite{blockchain} that is drastically different from usual
bank transactions deeply hidden from the public eye.

Since the data of bitcoin transaction 
network are open to public it is rather interesting to analyze the statistical properties
of this Bitcoin network (BCN). Among the first studies of BTN we quote
\cite{shamir} and \cite{ober,marcin} where the statistical properties of BCN have been studied
including the distribution of ingoing and outgoing transactions (links).
Thus it was shown that a distribution of links is characterized by a power law
\cite{ober,marcin} which is typical for complex scale-free networks \cite{dorogovtsev}.
Due to this it is clear that the methods of complex networks, such as the World Wide Web (WWW)
and Wikipedia, should find useful applications for the BCN analysis. 
In particular, one can mention in this context the important PageRank 
algorithm \cite{brin} which is at the foundation of the Google search 
engine \cite{meyer}.
Applications of this and related algorithms to various directed networks 
and related Google matrix are discussed in \cite{rmp2015}.
Previous studies of the world trade network \cite{wtn1,wtn2}
showed that for financial transactions or related trade of commodities 
it is useful to consider also the CheiRank probabilities 
for a network with inverted links \cite{linux}
and we will use this approach also here. In addition we 
analyze the spectrum of the Google matrix of BCN
using the powerful numerical approach of the Arnoldi algorithm
as described in \cite{ulamfrahm,integer_network,citation_network}.
We note that a possibility to use the PageRank probabilities for BCN was
briefly noted in \cite{mit}.

In our studies we use the bitcoin transaction  data 
collected by Ivan Brugere from the public block chain site \cite{blockchain} 
with all bitcoin transactions from the bitcoin birth in January 11th 2009 
till April 2013 \cite{ivan}.

The paper is composed as follows: In Section 2 we describe the main properties of BCN,
the Google matrix is constructed in Section 3,
the numerical methods of its analysis are described in 
Section 4, the spectrum and eigenvectors of $G$ matrix are 
analyzed in Sections 5 and 6, the Gini coefficient of BCN is determined
in Section 7 and the discussion is given in Section 8.

\section{Global BCN properties}

From the bitcoin transaction data \cite{ivan} of the period from 
the very beginning in January 11$th$ 2009 to April 10$th$ 2013,  
we construct the BCN and related Google matrix.
This weighted and directed network takes into account the sum of all transactions,
measured in units of bitcoin,
from one user to another during a given period of time.
The total number of transactions in this period is $N_t =28.140.756$.
The minimum transaction value is $10^{-8}$ (was $10^{-3}$) bitcoin for 
the period after (before) march 2010.

\begin{figure}[t]
\begin{center}
\includegraphics[width=0.46\textwidth]{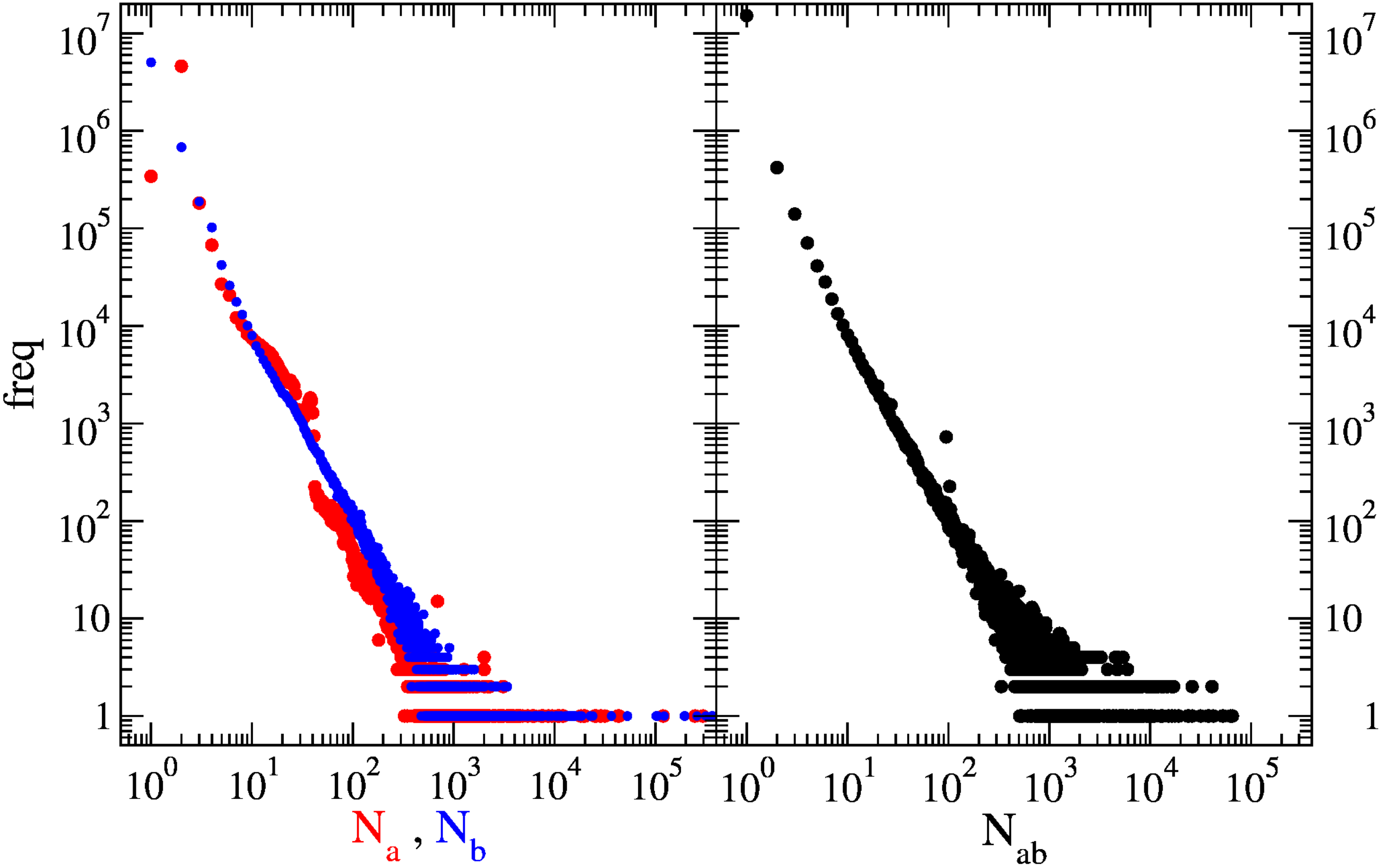}
\end{center}
\caption{\label{fig1}(color online) Frequency
histograms of BCN $N_f$ in the period from January 11$th$ 2009 to April 10$th$ 2013.
Left panel shows the frequency distribution 
$N_f$ of number of sellers (and buyers) transactions $N_a$ (and $N_b$).
Right panel shows the frequency of transactions with the same given partners $N_{a,b}$.
}
\end{figure}

\begin{figure}[t]
\begin{center}
\includegraphics[width=0.46\textwidth]{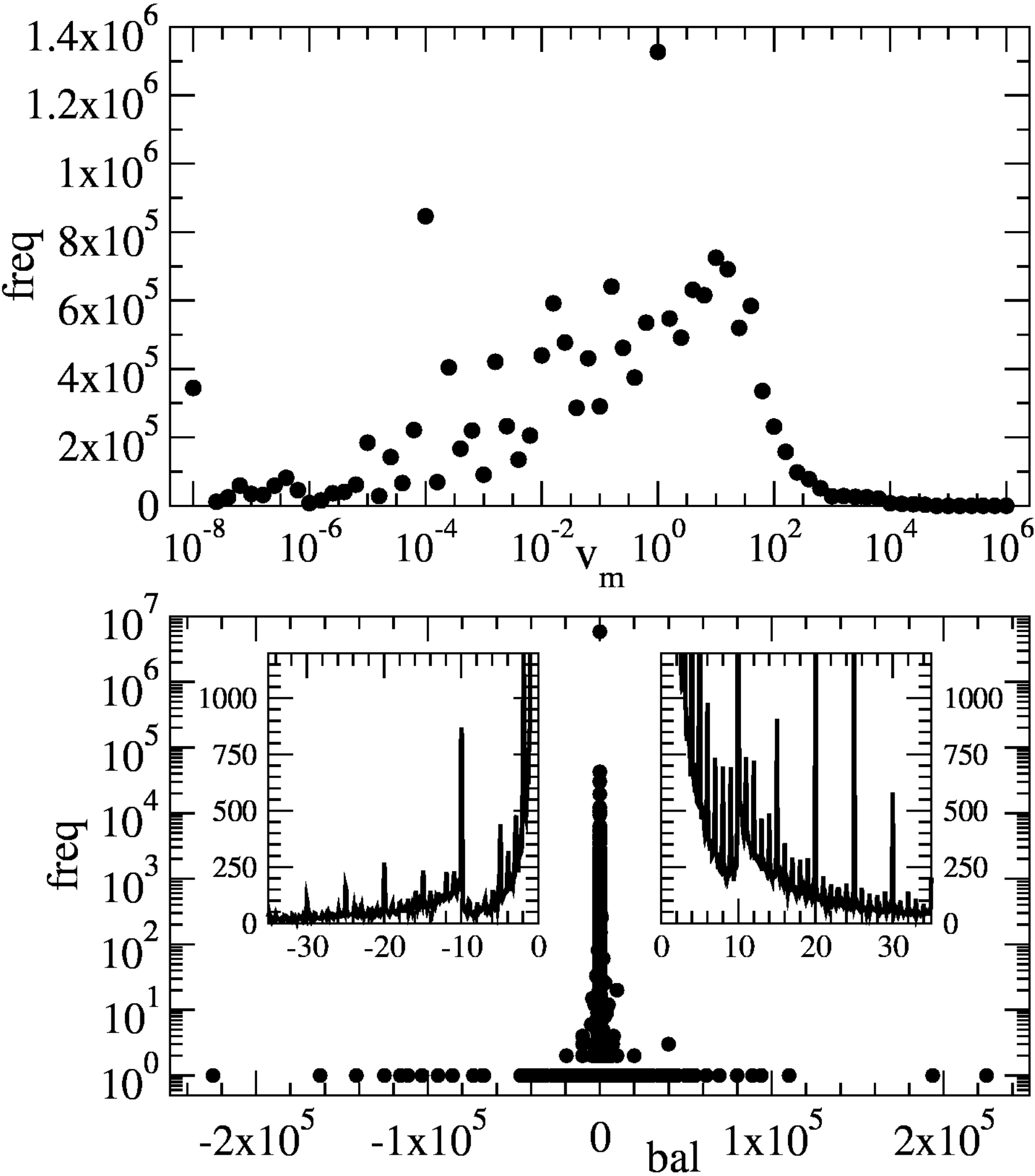}
\end{center}
\caption{\label{fig2}(color online)
Frequency histogram $N_f$  of bitcoin transaction volume $v_m$ measured in bitcoins on top panel
(histogram is equidistant in $\log_{10}v_m$ with a distance of $0.2$).
Bottom panel shows the frequency histogram $N_f$ of user balance $B_u$ defined as 
the difference between ingoing and outgoing transactions
in bitcoin units. 
Left and right insets show zoom in vicinity of zero balance for 
negative (left) and positive (right) values.
}
\end{figure}

\begin{figure}[t]
\begin{center}
\includegraphics[width=0.46\textwidth]{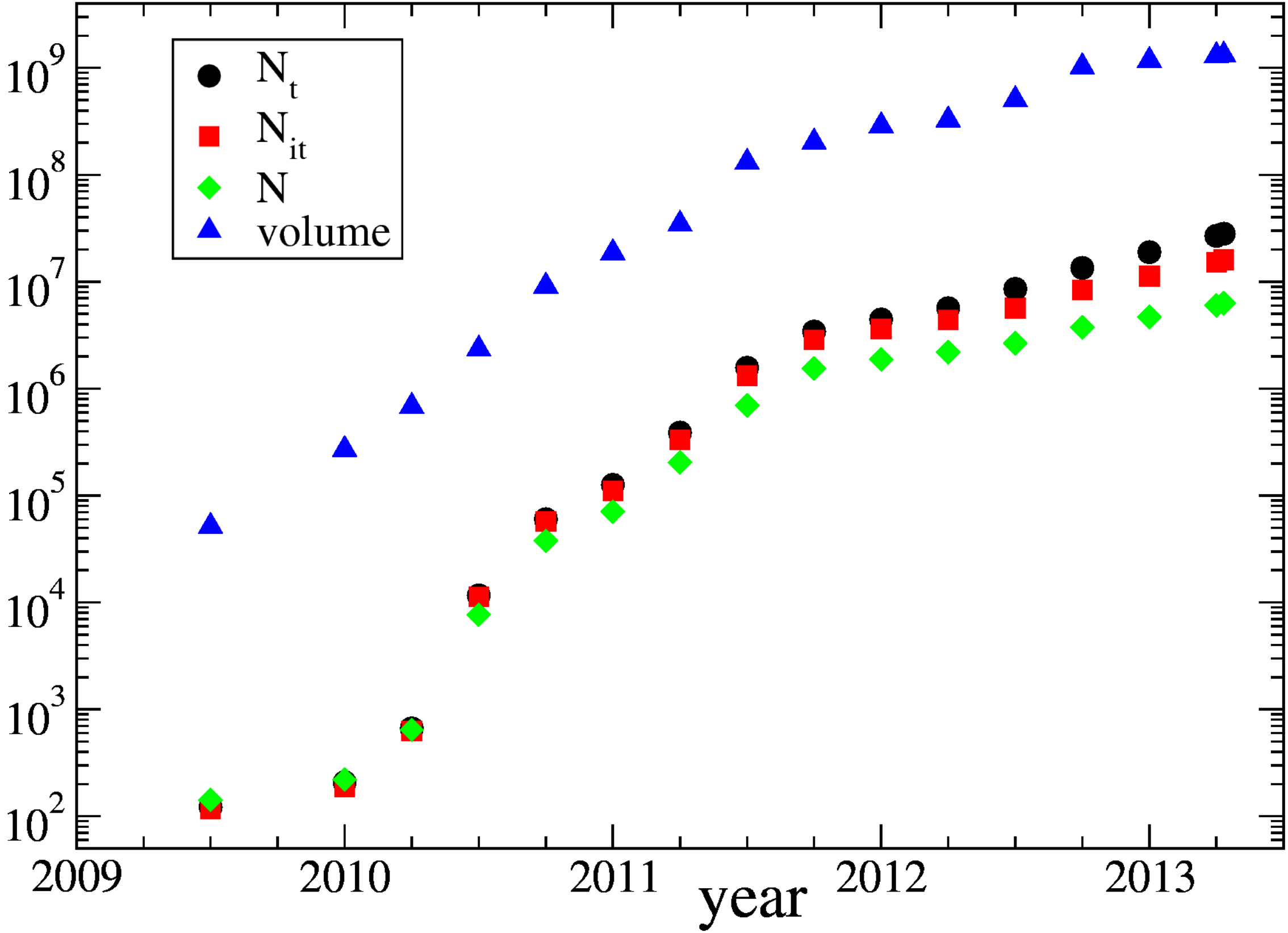}
\end{center}
\caption{\label{fig3}(color online)
Characteristic evolution of 
BCN with year quarters (halves in 2009) 
from 2009 to April 2013.
Time evolution is shown for  
number of transactions $N_t$ in a given quarter (black circles); 
number links $N_{it}$ of integrated transactions 
from the beginning till given quarter (red squares); 
number of nodes given by partners $N$ (green diamonds); and
total volume of bitcoins (blue triangles).
}
\end{figure}

The global statistical characteristics of transactions are shown in 
Figs.~\ref{fig1},~\ref{fig2},~\ref{fig3}.
Thus Fig.\ref{fig1} shows the frequency histogram $N_f(N_a), N_f(N_b)$ of BCN in this period, 
given the dependencies for outgoing links 
(or sellers $N_a$), ingoing links (or buyers $N_b$), and transactions of the same partners 
from $a$ to $b$ ($N_{a,b}$). The fit of the data 
is in a satisfactory agreement with an algebraic decay $N_f \propto 1/{N_a}^\beta$,
$N_f \propto 1/{N_b}^\beta$, $N_f \propto 1/{N_{a.b}}^\beta$ with 
$\beta = 2.1 \pm 0.1$, $\beta =1.8 \pm 0.1$, $\beta =2.2\pm0.1$ respectively.

Top panel of Fig.\ref{fig2} shows the histogram of bitcoin 
transaction volume $v_m$ for the whole period (2009-2013) measured in bitcoin.
It is visible that it has peaks in values of $10^{-8}, 10^{-4}$ and $1$.
At the same time there are also transactions with many bitcoins and $v_m$ as large as $834352.9$.
The balance of each user $B_u$  can be defined as the sum of all ingoing transactions minus
the outgoing ones measured in bitcoins. This balance $B_u$
is shown in the bottom panel of Fig.\ref{fig2}. 
For a majority of users the balance is close to zero
but in a few cases $B_u$ is strongly negative or positive.
There are 
also visible peaks at values $B_u=30, 25, 20, 10$.

In order to study BCN time evolution we divide the whole period of time 
in year quarters  from 
2009 to 2013 (we take only half of years in 2009 since the number of transactions is very small).
Some characteristic numbers of BCN are shown in Fig.\ref{fig3}. There is a significant growth
with time for the number of transactions $N_t$,
and the integrated number  of transactions $N_{it}$ (from the beginning till given quarter)
and the number of nodes $N$ (partners) for the same period of time.

At the next step we describe the construction of the Google matrix
from the bitcoin transactions described above.

\section{Construction of Google matrix of BCN}

In this work we use the notation ``BCYearQuarter'' (e.g. BC2010Q2)
for the different bitcoin networks, eventually with an additional ``*'' 
for the CheiRank case  \quad  \quad (e.g. BC2010Q2*). 
We consider 16 (or 32 including the $\;\;\;$ CheiRank cases) networks 
BC2009Q2, BC2009Q4 to $\;\;\;$ BC2013Q2 with network sizes $N$ and 
link numbers $N_\ell$ ranging from $N=142$ and $N_\ell=117$ (BC2009Q2) to 
$N=6297009$ and $N_\ell=16056427$ (BC2013Q2) with typical ratios 
$N/N_\ell$ between $1$ for the smallest networks and $2.5-3$ for 
the largest networks. For the whole period of all quarters 
we have the total $G$ matrix size $N=6297539$
with $N_\ell = 16056427$ links. The values of $N, N_\ell$ and total volume for all quarters
are given in Table~1. 

\begin{table}[h]
\begin{center}
\begin{tabular}{ | c | c | c| c | }
  \hline			
Network& $N$& $N_\ell$ & total volume \\
& & &(in bitcoins)\\ 
  \hline  
BC2009Q2& 142 & 117 &  51499\\
BC2009Q4& 220& 188&  269526\\
BC2010Q1& 645& 632& 681867\\
BC2010Q2& 7706& 11275& $2.33662\times10^6$\\  
BC2010Q3& 37818& 57437&  $9.0931\times10^6$\\
BC2010Q4& 70987& 111015& $1.86444\times10^7$\\
BC2011Q1& 204398& 333268& $3.44654\times10^7$\\
BC2011Q2& 697401& 1328505& $1.30747\times10^8$\\
BC2011Q3& 1547349& 2857232&  $2.0177\times10^8$\\
BC2011Q4& 1885400& 3635927&  $2.87714\times10^8$\\
BC2012Q1& 2186598& 4395611& $3.2546\times10^8$\\
BC2012Q2&  2645532& 5655802& $5.04581\times10^8$\\
BC2012Q3&  3742691& 8381654& $1.02381\times10^9$\\
BC2012Q4&  4672122& 11258315& $1.17078\times10^9$\\
BC2013Q1&  5998239& 15205087& $1.29944\times10^9$\\
BC2013Q2&  6297009& 16056427& $1.31479\times10^9$\\
  \hline  
\end{tabular}
\caption{\label{table1} Size ($N$), number of links ($N_\ell$) and total volume of used networks.}
\end{center}
\end{table}

As usual we write the matrix associated to such a network 
as \cite{rmp2015,citation_network}:
\begin{equation}
\label{eqSdefine}
S=S_0+\frac{1}{N}\,e\,d^T
\end{equation}
where $e^T=(1,\ldots,N)$ is the (transpose of the) uniform vector 
with unit entries, $d$ is the dangling vector with unit entries $d_l=1$ 
if $l$ corresponds to an empty column of $S_0$ and $d_l=0$ for the other 
columns. The elements $(S_0)_{lk}$ of the matrix $S_0$ correspond to the value 
of the bitcoin transaction from a node $k$ to another node $l$ 
normalized by the total 
value of transactions from the node $k$ to all nodes. 
A similar construction of $S_0$ is used for the world trade network \cite{wtn1}.
For the CheiRank case \cite{linux} the direction of the transaction is inverted 
in this scheme, i.e. $(S_0^*)_{lk}$ corresponds to the value 
of the bitcoin transaction from the node $l$ to $k$ normalized 
by the total value of transactions from all nodes to the node $k$. 
According to our raw data the bitcoin transactions up to 2010Q2 were 
done in units of $10^{-3}$ bitcoins and afterwards in units of 
$10^{-8}$ bitcoins. Therefore the raw transaction values and also the 
resulting (column sum normalized) entries of the matrix $S_0$ are 
rational numbers. For computations using normal precision 
numbers (i.e. standard double precision with a mantissa of 52 bits) 
these rational numbers can simply be replaced by the closest floating point 
number. However, for high precision 
computations using the library GMP \cite{gmplib}, the precise rational 
values were kept as long as possible and, only when necessary, rounded to high 
precision floating point values with their maximal precision. 

For the purpose of PageRank computations we also consider the Google matrix 
with damping factor $\alpha$ given by:
\begin{equation}
\label{eqGdefine}
G=\alpha S+(1-\alpha)\frac{1}{N}\,e\,e^T
\end{equation}
where we use $\alpha=0.85$ corresponding to its typical choice \cite{brin,meyer,rmp2015}.
For the network with inverted direction of transactions, corresponding to the CheiRank case,
we have $G^* = \alpha S^* + (1-\alpha)\frac{1}{N}\,e\,e^T$.

\begin{figure}[t]
\begin{center}
\includegraphics[width=0.46\textwidth]{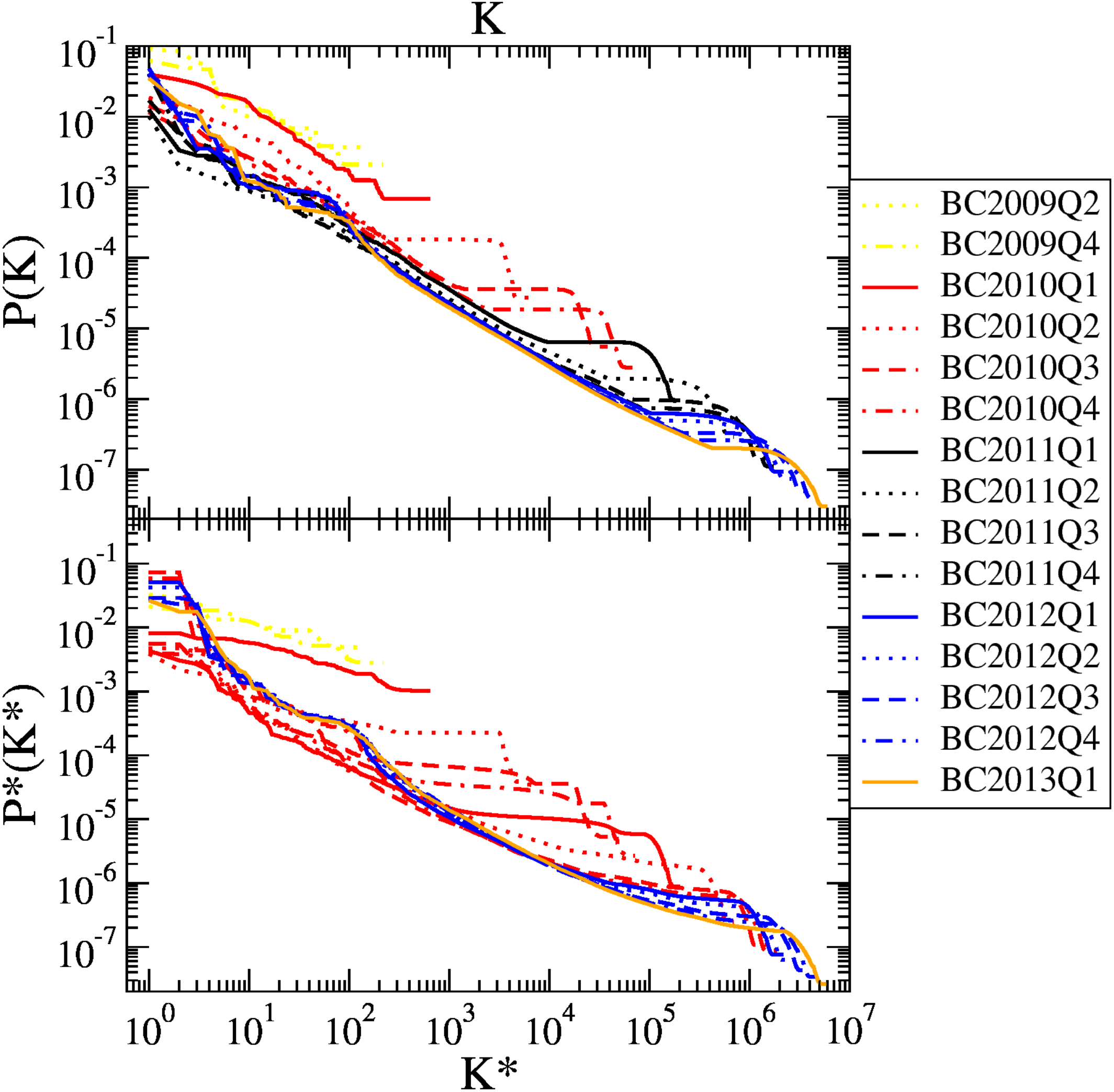}
\end{center}
\caption{\label{fig4}(color online)
PageRank and CheiRank distributions ordered by indices 
$K$ and $K^*$ on top and bottom panel respectively.
The bitcoin networks are taken by quarters of years (halves in the case of 2009) for 2009 (yellow),
2010 (red), 2011 (black), 2012 (blue) and 2013 (orange) with lines corresponding to Q1 (solid line),
Q2 (dotted line), Q3 (dashed line) and Q4 (dot-dashed line).
}
\end{figure}

The right eigenvectors $\psi_m$ of $G$ are determined by the equation
$ \sum_{j'}  G_{jj'} \psi_i(j') = \lambda_i \psi_i(j)$ with eigenvalues $\lambda_i$.
At $\alpha <1$ the largest eigenvalue is $\lambda=1$ and the corresponding eigenvector
has only positive component which have (for WWW networks) 
the meaning of probabilities $P(j)$ ($\sum_j P(j) =1$)
to find a random surfer on a node $j$ \cite{meyer}. We can order 
all nodes in the order of monotonic decrease of probability $P(K)$
with maximal probability at the PageRank index $K=1$ and then at $K=2,3...$.
In a similar way for the CheiRank case of $G^*$ we obtain the CheiRank 
vector at $\lambda=1$ with CheiRank probability $P^*(K^*)$ being maximal
at the CheiRank index   $K^*=1$ and then at $K^*=2,3...$.
The PageRank vector is efficiently determined by the power iteration algorithm
\cite{brin,meyer}.

The dependencies of the PageRank  $P(K)$ and CheiRank  $P^*(K^*)$ probabilities 
on their indices $K, K^*$
are shown in Fig.~\ref{fig4} for various quarters of BCN.
We see that the distributions become stabilized at last quarters when the network 
size becomes larger reaching its steady-state regime.
Thus for BC2013Q1 we find that the probability approximately decays 
in a power law with $P \propto 1/K^\nu, P^* \propto 1/{K^*}^\nu$
with $\nu = 0.86 \pm 0.06$, $ \nu =0.73 \pm 0.04$ respectively
(the fit is done for the range $10 < K,K^* < 10^5$).
The value of $\nu$ is similar to the values found for other directed networks
(see e.g. \cite{rmp2015,wtn1,wtn2}) but we note that this is only
an approximate description of the numerically found behavior
(see detailed discussion of algebraic decay for WWW networks in \cite{vigna}).

\section{Numerical methods for BCN Google matrix diagonalization}

We describe here the various skillful numerical methods used for diagonalization of
$G$ and $G^*$. Their use had been required due to heavy numerical problems for
accurate computation of the eigenvalues 
of these matrices and related eigenvectors.

First we introduce the concept of invariant isolated subsets
(for more details we refer to \cite{fgsjphysa}).
These subsets are invariant with respect to applications of $S$. 
The remaining nodes not belonging to an invariant subset 
(below a certain maximum size, e.g. 10\% of the network size) 
form the wholly connected {\it core space}. 
The practical computation of these subsets can be efficiently implemented in a 
computer program \cite{fgsjphysa}, eventually merging 
subspaces with common members, which provides a sequence of 
disjoint subspaces invariant by applications of $S$. Therefore we obtain 
a subdivision of the network nodes in $N_c$ 
core space nodes and $N_s$ subspace nodes 
(belonging to at least one of the invariant subsets) corresponding 
to the block triangular structure of the matrix $S$:
\begin{equation}
\label{eq3}
S=\left(\begin{array}{cc}
S_{ss} & S_{sc}  \\
0 & S_{cc}\\
\end{array}\right)\;\ .
\end{equation}
Here $S_{ss}$ is composed of many small diagonal blocks for 
each invariant subspace and whose eigenvalues can be efficiently obtained 
by direct (``exact'') numerical diagonalization.

We have computed for the networks up to BC2011Q4 (with 
$N=1884918$ and $N_\ell=3635927$) (a part of) the complex eigenvalue spectrum 
of the matrix $S$ (i.e. $G(\alpha)$ for $\alpha=1$) with eigenvalues closest 
to the unit circle. For this we employed 
basically the method of Refs. \cite{fgsjphysa,wikispectrum} 
based on (\ref{eq3}) to compute exactly the eigenvalues associated to 
the invariant subsets, typically 
a very modest number. For each invariant subspace there is at least 
one unit eigenvalue $\lambda=1$ which is therefore possibly degenerate 
(in case of several invariant subspaces). 
The remaining eigenvalues associated to the main 
core space (with $|\lambda|<1$) are obtained by the Arnoldi 
method \cite{arnoldibook,ulamfrahm} with Arnoldi dimensions up to $n_A=16000$. 
This requires for the network BC2011Q4 a machine with 256 GB (using standard 
double precision numbers). 

For the larger networks (BC2012Q1 and later) 
it would  be necessary to increase the available memory or to reduce the 
value of $n_A$. However, it turns out that the density of eigenvalues close 
to the unit circle is so high that a significant reduction of $n_A$ does 
not allow to obtain (even a small number) of reliable core space eigenvalues. 
This situation is quite different from other networks such as certain 
university networks \cite{fgsjphysa} or Wikipedia \cite{wikispectrum} where 
it was easier to access numerically a reasonable number of the top 
core space spectrum of the matrix $S$. 
Furthermore for the cases up to BC2011Q4 we also computed at least 
20 eigenvectors of 20 selected (core space) eigenvalues close to the unit 
circle such that roughly $\lambda_j\approx |\lambda_j|\exp(i2\pi j/19)$ for 
$j=0,\ldots,19$ and $|\lambda_j|\approx 1$. 
 
For the smallest bitcoin networks BC2009Q2, $\;\;\;\;$  \qquad BC2009Q4 
and BC2010Q1 
with $N\le 645$ the core space eigenvalue spectrum is actually easily 
accessible by direct diagonalization or full Arnoldi diagonalization 
(with some subtle effects for the small eigenvalues requiring high precision 
computations). 

The four networks BC2010Q2 and BC2010Q2* $\;\;\;\;$ (BC2010Q3 and BC2010Q3*)
play a somewhat special role in our studies since on 
one hand they are sufficiently small with $N=7706$ (or $N=37818$) to allow 
(at least in theory) to compute all (or nearly all) non-zero eigenvalues 
and on the other hand they are still sufficiently large to have an 
interesting spectrum, comparable to the spectra of the larger networks, 
especially with a strong concentration of the majority of (non-vanishing) 
eigenvalues close to the unit circle. 

However, it turns out that the two cases of BC2010Q2 and BC2010Q2* 
suffer from a serious numerical 
problem similar to the citation network of Physical Review 
\cite{citation_network}. Using both direct diagonalization 
(i.e. using Householder transformations to transform the initial matrix to 
Hessenberg form and final diagonalization of the latter by the QR algorithm 
with implicit double shift) and full Arnoldi diagonalization (choosing 
a sufficiently large value of $n_A$ and QR algorithm to diagonalize the 
Arnoldi matrix which is also of Hessenberg form) with normal precision 
floating point numbers we find that there are several ``rings'' of eigenvalues 
close to the unit circle. The outer two rings seem to contain reliable and 
correct eigenvalues but already the third ring with $|\lambda|\approx 0.94$ 
and all rings below are numerically completely unreliable since the 
corresponding 
eigenvalues change completely between the two methods and also different 
implementations of them (i.e. applying a permutation in the network nodes but 
keeping the same network structure, choosing different ordering in the 
summation when computing the scalar products for the Arnoldi method, using 
slightly different but mathematical equivalent implementations of the 
QR algorithm, using different runs with  parallelization which amounts 
to different rounding errors for the sums in the scalar products etc.). 
Therefore we conclude that eigenvalues with $|\lambda|< 0.95$ are 
numerically incorrect as long as we use methods based on normal precision 
numbers. 

This situation is quite similar to the (nearly) triangular 
citation network of Physical review \cite{citation_network} 
where eigenvalues with $|\lambda|<0.4-0.5$ are numerically wrong. 
The reason of this behavior is due to large Jordan blocks for the 
highly degenerate zero eigenvalue producing numerically artificial rings of 
incorrect eigenvalues in the complex plane with radius 
$r\sim \varepsilon^{1/d}$ \cite{integer_network,citation_network} with 
$\varepsilon$ being the machine precision (i.e. $\varepsilon=10^{-16}$ 
for simple double precision numbers or $\varepsilon=2^{-p}$ for high 
precision numbers with $p$ binary digits) and $d\gg 1$ being the dimension 
of the Jordan block. The bitcoin networks do not have the (near) triangular 
structure, responsible for this problem in \cite{citation_network}, 
but the low ratio of $N_\ell/N\approx 1.5$, reducing considerably 
the number of non-zero matrix elements in $S_0$, also creates large 
Jordan subspaces and here the 
effect is even worse as compared to Ref. \cite{citation_network}. 

To solve this problem and obtain final reliable eigenvalues with 
precision $10^{-15}$, we implemented all steps of the numerical 
diagonalization methods: the computation of the Arnoldi 
decomposition, reduction of an arbitrary matrix to Hessenberg form 
using Householder transformations, final diagonalization of Hessenberg 
matrices by the QR algorithm, with high precision floating point 
numbers using the GMP library \cite{gmplib}. (In Ref. \cite{citation_network} 
only the computation of the Arnoldi decomposition was implemented 
with the GMP library.). 

For the two networks BC2010Q2 and BC2010Q2* we have been able to push the 
direct  high precision diagonalization 
(Householder transformation to Hessenberg form and QR algorithm) with 
different precision up to $p=4096$ binary digits confirming the scaling 
$r\approx 2^{-p/d}$ for the radius of incorrect eigenvalues induced by 
large Jordan blocks. For $p=4096$ we find a maximal radius $r\approx 0.01$ 
corresponding to a value of $d\approx 616$ for the dimension of the 
corresponding Jordan block. In normal precision (with $p=52$) the 
same value of $d$ corresponds to a radius $\approx 0.94$ confirming 
exactly the observations of the initial normal precision results. 

The direct diagonalization in high precision is however quite expensive 
in both computation time and memory requirement. In this context the 
(high precision) Arnoldi method is more efficient since it automatically 
breaks off 
when it has explored an $S$-invariant subspace which is detected by 
a vanishing or very small coupling matrix element in the Arnoldi matrix 
at some value of $n_A$ 
(see Refs. \cite{ulamfrahm,citation_network} for 
more details on this point). If we assume that the initial vector 
(which we chose either uniform or random with two different realizations) 
contains contributions from all eigenvectors associated to non-vanishing 
core space eigenvalues the method will, at least in theory, produce the 
complete spectrum of these eigenvalues using a considerably reduced subspace 
for the final (QR-) diagonalization. Here we have chosen a break off limit 
of $\epsilon=2^{-p/2}$ (with $p$ being the precision number of binary digits) 
for the final 
coupling matrix element which scales to zero with increasing precision 
but is still much larger than the computation precision ($2^{-p}$) allowing 
to take into account the subtle effects due to the Jordan blocks creating 
numerical errors on a scale much larger than the computation precision. 
In this case we obtain a reduced dimension of about 2000-3000 (depending 
on the choice of random or uniform initial vectors and on both cases 
of BC2010Q2 or BC2010Q2*) instead of 7706. Here the Arnoldi method 
with a precision of $p=8192$ binary digits (which is considerably less 
expensive than the direct diagonalization with $p=4096$) or even only $p=3072$ 
(for the case of BC2010Q2* with uniform initial vector) allows to 
obtain the complete spectra of non-vanishing eigenvalues for these 
two networks. The remaining small rings of 
numerical incorrect Jordan block induced eigenvalues can be 
easily removed from the correct eigenvalues by comparing the spectra 
obtained by different initial vectors. 

We also employed (with some suitable technical modifications which we omit 
here) the rational 
interpolation method which we developed in Ref. \cite{citation_network}. 
This method is also based on high precision computations to determine 
the zeros of a certain rational function which are the core space 
eigenvalues satisfying the condition $d^T\,\psi\neq 0$ for the 
corresponding eigenvector $\psi$ and the above introduced dangling vector 
$d$. It turns out that for the two networks BC2010Q2 and BC2010Q2* 
all non-vanishing core space eigenvalues satisfy this condition but 
for the other two networks BC2010Q3 and BC2010Q3* there a few core space 
eigenvalues with $d^T\,\psi=0$ which we determined separately 
by a method described in Ref. \cite{citation_network} exploiting that 
they are degenerate subspace eigenvalues of the matrix $S_0$ (which are 
different from the subspace eigenvalues of $S$ which we also computed). 

The rational interpolation method is highly effective with very modest 
memory requirements and the possibility to use partial low-precision 
spectra to accelerate the computation of the zeros to obtain recursively 
higher precision spectra. 
Here we obtained for BC2010Q2 and BC2010Q2* 
precise and complete spectra for $p=6144$ but we also performed 
confirmation runs up to $p=12288$. The results of this method 
confirm exactly the numerical values 
(with accuracy of $10^{-15}$ for all of the final eigenvalues) and the precise 
number of non-vanishing core space eigenvalues already obtained by the 
high precision Arnoldi method. We mention that the eigenvalues of the 
direct diagonalization 
correspond numerically with the same accuracy to these results (after 
removal of the numerically incorrect Jordan induced eigenvalues) but for 
$p=4096$ this method misses a small number (about $3-4$) of the smallest 
non-vanishing core space eigenvalues (with $|\lambda|\sim 5\times 10^{-3}$). 

\section{Spectrum of BCN Google matrix}

We present here the main results obtained for the spectrum and some eigenvectors
of $G$ and $G^*$ by the numerical methods described above.

\begin{figure}[h]
\begin{center}
\includegraphics[width=0.48\textwidth]{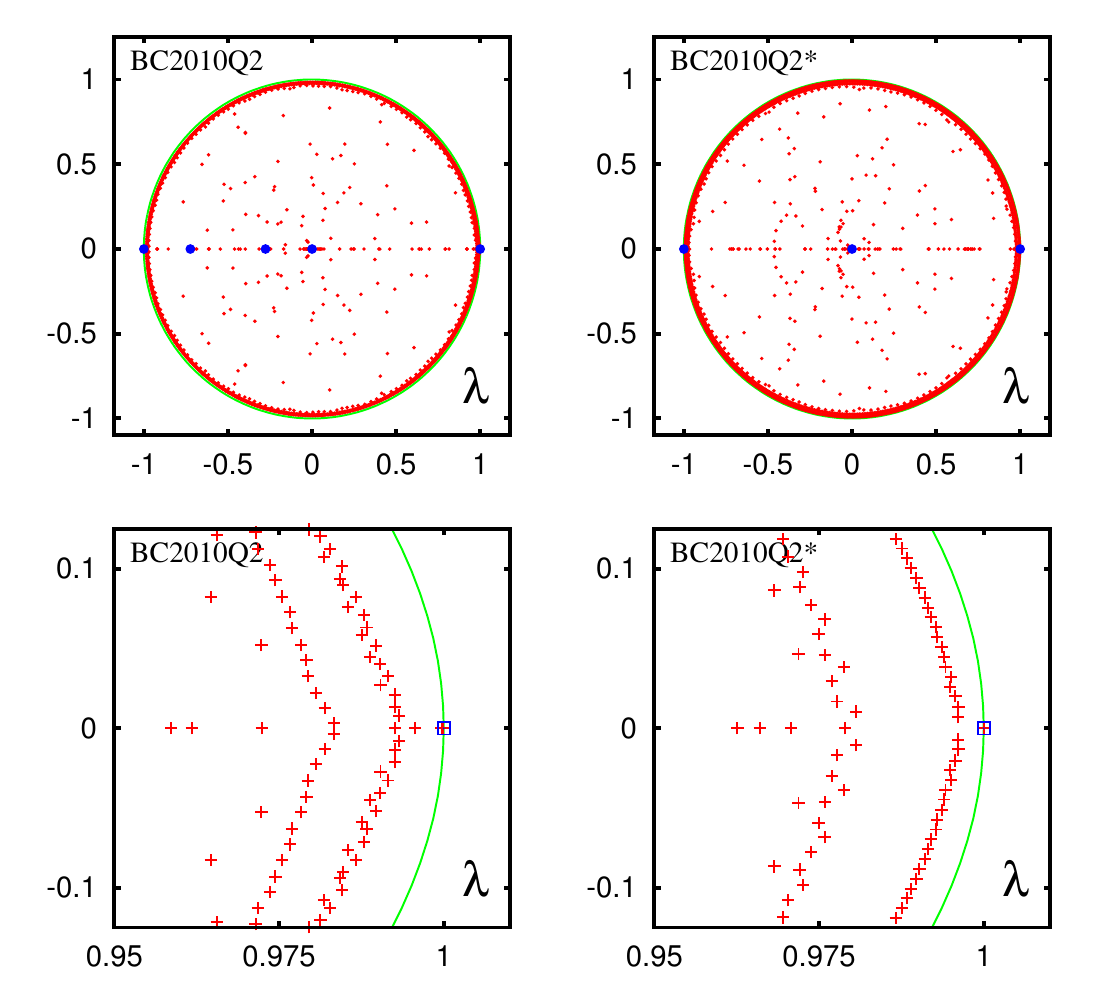}
\caption{(Color online)
Complex eigenvalue spectrum of the Google matrix associated 
to the network BC2010Q2 (BC2010Q2*) in left (right) panels. 
Shown are the full spectrum in top panels or a zoomed representation 
for the region $\lambda\approx 1$ in bottom panels. 
The red dots (crosses) are core space eigenvalues obtained 
by high precision Arnoldi computations and also the rational interpolation 
method and the blue thick dots (square boxes) are 
invariant subspace eigenvalues 
obtained by the normal/high precision Arnoldi method or direct diagonalization.
The green line (if visible and not hidden by the red dots) is the unit circle. 
In top panels the apparent ``red circle'' corresponds in reality to 
a high density of individual red dots for the complex core space eigenvalues 
whose structure is better visible in the zoomed representation
in bottom panels. 
The top core space eigenvalues (red crosses) which are very close to the top 
sub space eigenvalue at $\lambda=1$ (blue square box) are 
$0.99990029706715$ and $0.999678494064214$ (or $0.999998157039589$)
for BC2010Q2 (BC2010Q2*). The 3rd top core space eigenvalue 
$0.995663863983884$ for BC2010Q2 is already clearly outside the 
blue square box. More details for the computation method and 
the subspace eigenvalues are given in the main text. 
}
\label{fig5}
\end{center}
\end{figure}

For the two networks BC2010Q2 and BC2010Q2*, with a full network size of 
$N=7706$, we find that there are exactly $N_c=1967$ 
($N_c=1984$) non-vanishing core space eigenvalues and $N_s=15$ ($N_s=2$)
non-vanishing subspace eigenvalues (of $S$) for BC2010Q2 (or BC2010Q2*) 
with the complete and numerically accurate spectra shown in Fig.~\ref{fig5}. 
The main two outer rings close to the unit circle (with $|\lambda|>0.97$) 
contain 1626 (1621) core space eigenvalues which is more than 
$80\%$ of the spectrum (of non-vanishing eigenvalues). 
The non-vanishing subspace eigenvalues $\lambda$, also shown in 
the same figure, and 
their multiplicities $m$ are $m=7$ ($\lambda=1$), $m=6$ ($\lambda=-1$), 
$m=1$ ($\lambda=-0.723606797749979$ and $\lambda=-0.276393202250021$) 
for BC2010Q2 and $m=1$ ($\lambda=\pm 1$) for BC2010Q2*. All other eigenvalues 
(about $\sim 5700$) are zero and correspond to Jordan subspaces with 
potentially rather large dimensions being responsible for the numerical 
problems when limiting the computations to normal floating point precision. 

\begin{figure}[h]
\begin{center}
\includegraphics[width=0.48\textwidth]{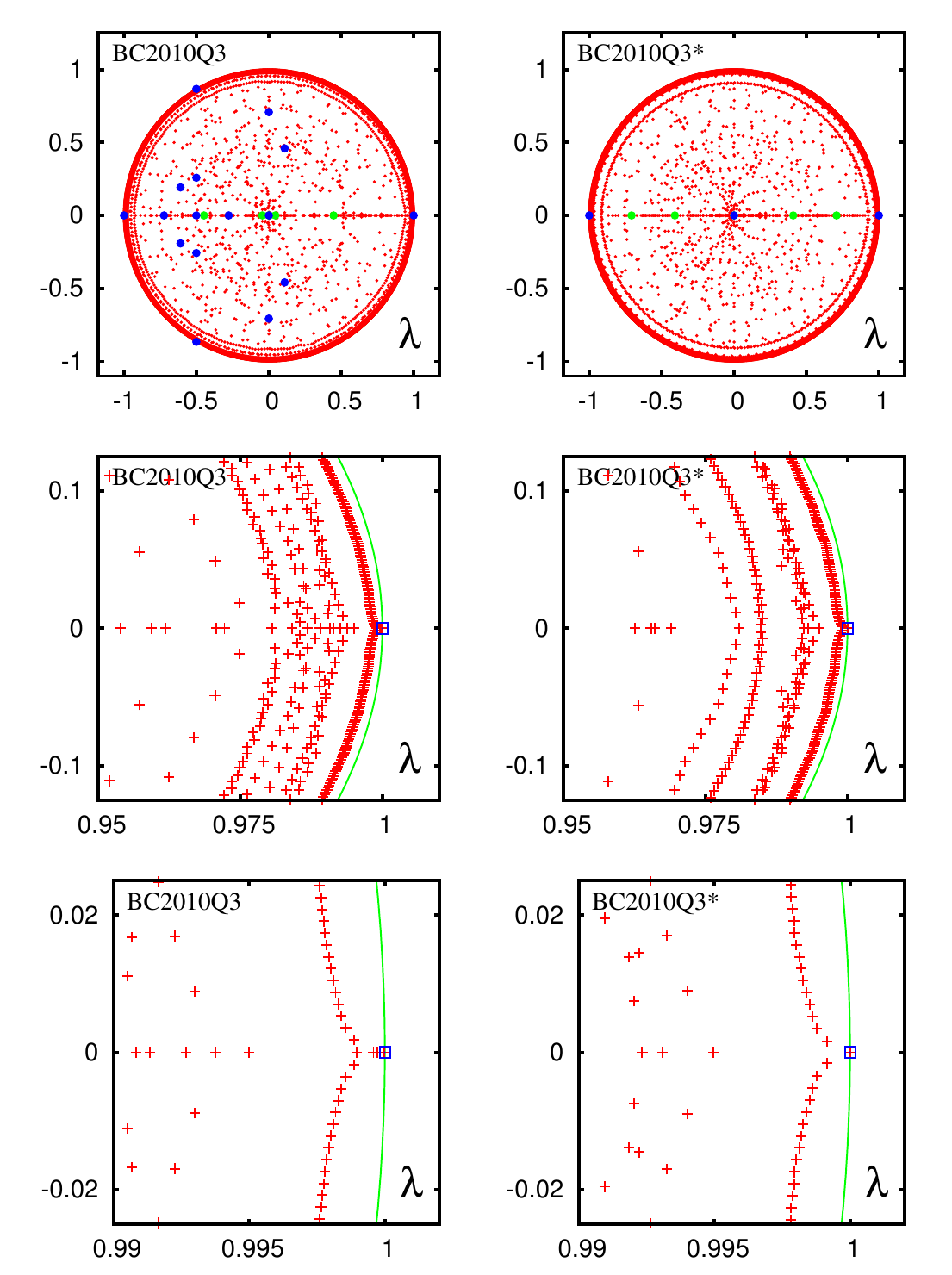}
\caption{(Color online)
Complex eigenvalue spectrum of the Google matrix associated 
to the network BC2010Q3 (BC2010Q3*) in left (right) panels. 
Shown are the full spectra in top panels and two zoomed representations 
for the region $\lambda\approx 1$ in center and bottom panels. 
The red dots (crosses) are core space eigenvalues obtained by the 
rational interpolation method in high precision, 
the blue thick dots (square boxes) are 
invariant subspace eigenvalues 
obtained by the normal/high precision Arnoldi method or direct diagonalization 
and the thick green dots in top panels correspond to degenerate subspace 
eigenvalues of $S_0$ which are also core space eigenvalues of $S$ 
and not necessarily found by the rational interpolation method 
(see Ref. \cite{citation_network} for explanations).
There are 2 (is 1) top core space eigenvalue(s) 
(red cross(es)) very close to the top sub space eigenvalue at $\lambda=1$ 
i.e. nearly or completely inside the blue square box (in bottom panels) 
and the 1st top core space eigenvalue is $0.999968720409915$ 
($0.99999983940032$) for BC2010Q3 (BC2010Q3*).}
\label{fig6}
\end{center}
\end{figure}

For the two networks BC2010Q3 and BC2010Q3*, with a full network size of 
$N=37818$, the numerical problems due to Jordan blocks for the zero 
eigenvalue are less severe but still present. Here the normal 
precision Arnoldi method allows to compute about 7800-7900 reliable 
eigenvalues within an error of $10^{-6}$ and which are rather strongly 
localized close to the boundary circle (if one tries larger values 
of $n_A$ one obtains only numerical incorrect eigenvalues). 
Here the high precision Arnoldi method is strongly limited due to 
memory requirements and it is not possible to go beyond a precision 
of $p=512$ which produces about 500-700 additional reliable eigenvalues and 
the resulting spectra are still quite concentrated close to the boundary 
circle. 
However, the rational interpolation method still works very 
well due to its high efficiency. It turns that at a binary precision 
of $p=30720$ using about 18400 support points (for the rational 
interpolation scheme) this method produces $N_c=9192$ ($N_c=9145$) 
non-vanishing core 
space eigenvalues (including 4 pairs of doubly degenerate eigenvalues in 
both cases). However, without going into technical details, 
our results indicate that these numbers may still 
increase very slightly when increasing the precision and also the number of 
support points but we are confident that for both networks BC2010Q3 and 
BC2010Q3* there are about $N_c\approx 9200$ non-vanishing core space 
eigenvalues which is about 25\% of the full network size (a similar ratio 
we already found for BC2010Q2 and BC2010Q2*). 
The additional 1300-1400 eigenvalues with respect to the spectra obtained by 
the normal precision Arnoldi method fill out rather uniformly the inner part 
of the complex unit circle as can be seen in Fig.~\ref{fig6}. 

Furthermore for BC2010Q3 (BC2010Q3*) there also 
$N_s=56$ ($N_s=2$) subspace eigenvalues for $S$ (blue $\;\;\;$
dots/squares in Fig.~\ref{fig6}). Here some eigenvalues 
are on the unit circle with $|\lambda|=1$ and 
degeneracy $m=23$ ($m=1$) for $\lambda=1$, $m=17$ ($m=1$) for $\lambda=-1$
and $m=2$ ($m=0$) for $\lambda=(-1\pm i\sqrt{3})/2$. 
In both cases there also a few core space eigenvalues (given as 
degenerate subspace eigenvalues of $S_0$, green dots) which were determined 
by another method \cite{citation_network} 
since they are not necessarily found by the rational interpolation method. 
About $8000$ reliable eigenvalues are found by 
the normal precision Arnoldi method correspond to the 4-5 rings of eigenvalues 
close to the unit circle and visible in the center panels of Fig.~\ref{fig6}.

We mention that the high precision variants of the three methods are 
also useful to compute the full spectra for the three smaller networks 
(up to BC2010Q1 with $N=645$) and also for the invariant subspace spectra 
(for nearly all bitcoin networks) since they allow to remove in a reliable way 
a certain number of numerical incorrect eigenvalues below $10^{-3}$ obtained 
by the normal precision computations. For these cases the computation times 
are negligible and the required precision is rather modest (typically between 
$p=256$ and $p=1024$). Here the number of non-vanishing 
core space eigenvalues $N_c$ and subspace eigenvalues $N_s$ are given 
by $N_c=4$ ($6$) and $N_s=2$ ($0$) for BC2009Q2 (or BC2009Q2*) with 
$N=142$, $N_c=13$ ($15$) and $N_s=4$ ($2$) for BC2009Q4 (or BC2009Q4*) with 
$N=220$ and $N_c=26$ ($30$) and $N_s=6$ ($2$) for BC2010Q1 (or BC2010Q1*) 
with $N=645$. The subspace eigenvalues are always $\pm 1$ (except 
for BC2009Q2* where $N_s=0$ and there are no subspace eigenvalues) 
eventually with double (or triple) degeneracy if $N_s=4$ (or $N_s=6$). 
Clearly in all these cases the number of the non-vanishing 
core space and subspace eigenvalues constitutes only a small fraction of 
the spectrum with all other eigenvalues being zero corresponding 
to certain Jordan subspaces. 

\begin{figure}[h]
\begin{center}
\includegraphics[width=0.48\textwidth]{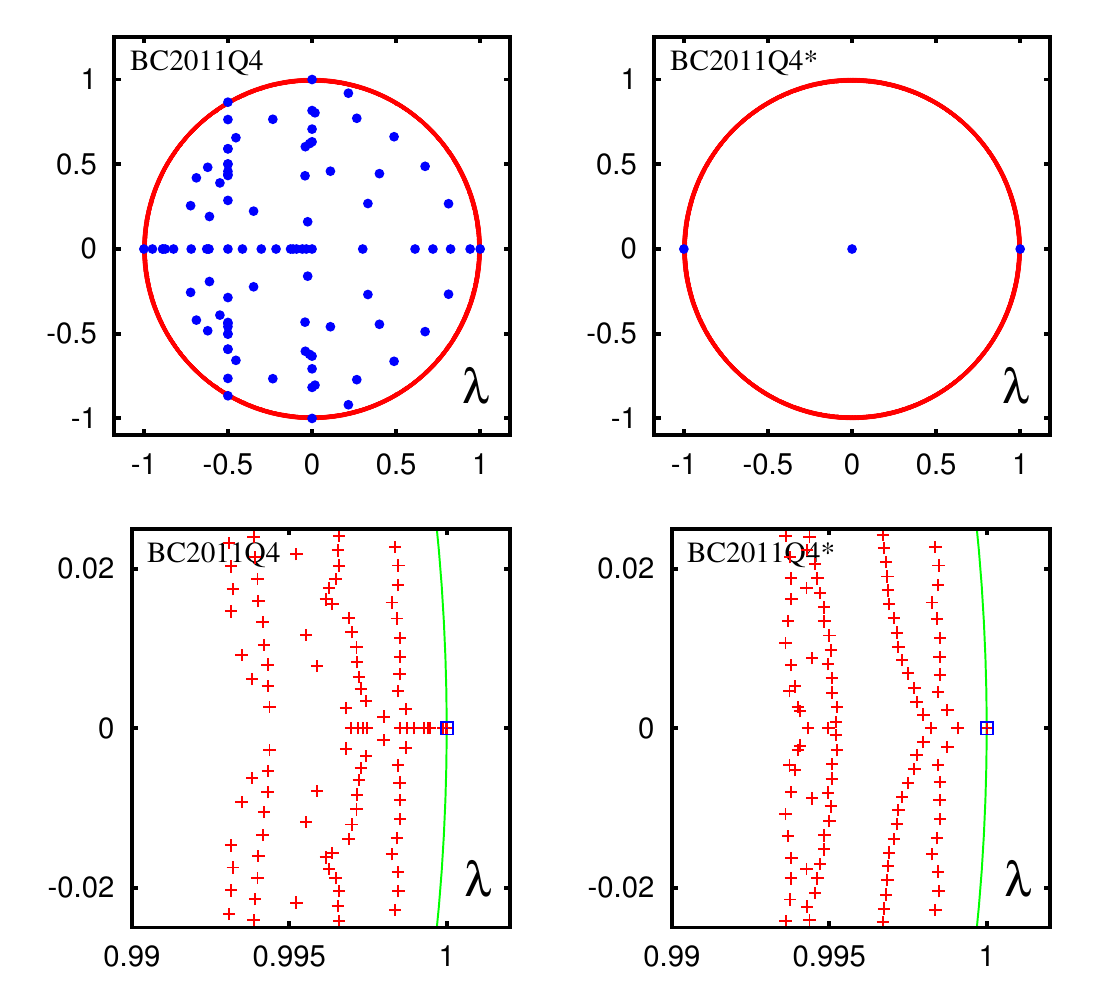}
\caption{(Color online)
Complex eigenvalue spectrum of the Google matrix associated 
to the network BC2011Q4 (BC2011Q4*) in left (right) panels. 
Shown are about 12000 ``reliable'' top eigenvalues obtained 
by the Arnoldi method in normal precision 
with $n_A=16000$ in top panels (red dots/red circle) or a zoomed 
representation (red crosses) for the region $\lambda\approx 1$ 
in bottom panels. The blue thick dots (square boxes) are 
invariant subspace eigenvalues 
obtained by the normal/high precision Arnoldi method or direct diagonalization.
The green line (if visible and not hidden by the red dots) is the unit circle. 
In top panels the apparent ``red circle'' corresponds in reality to 
a high density of individual red dots for the complex core space eigenvalues 
whose structure is better visible in the zoomed representation. 
There are 8 (is 1) top core space eigenvalue(s) 
(red cross(es)) very close to the top sub space eigenvalue at $\lambda=1$ 
i.e. nearly or completely inside the blue square box 
and the 1st top core space eigenvalue is $0.99999999417$ ($0.99999996048$) for 
BC2011Q4 (BC2011Q4*).
}
\label{fig7}
\end{center}
\end{figure}

For the larger networks (between BC2010Q4 with $N=70987$ and BC2011Q4 with 
$N=1884918$) we applied 
the normal precision Arnoldi method with $n_A=16000$. However, in view 
of the numerical problems visible for BC2010Q2/3, we performed different runs 
with slightly different implementations (e.g. different summation order 
for the scalar product in the Arnoldi method) leading to different rounding 
errors and verified how many eigenvalues were numerically identical with 
an error below $10^{-6}$. For the two cases BC2011Q4 and BC2011Q4* with 
$N=1884918$ and $n_A=16000$ we obtain about 12000 numerically reliable 
core space eigenvalues shown in Fig.~\ref{fig7} and which are all very close 
to the unit circle with $|\lambda|>0.99$. 
Fig.~\ref{fig7} also shows the subspace eigenvalues with 
$N_s=332$ ($2$) for BC2011Q4 (BC2011Q4*). The subspace spectrum 
of BC2011Q4 contains 242 eigenvalues on the unit circle with 
$|\lambda|=1$ which are $\lambda=1$ (degeneracy $m=127$), 
$\lambda=\pm i$ (both with $m=1$), 
$\lambda=(-1\pm i\sqrt{3})/2$ (both with $m=3$) and $\lambda=-1$ ($m=107$). 
The remaining 90 subspace eigenvalues with $0<|\lambda|<1$ are also 
visible in Fig.~\ref{fig7}. Here only one eigenvalue at $\lambda=-1/2$ 
has a double degeneracy. 
The subspace spectrum of BC2011Q4* contains only the two (non-vanishing) 
eigenvalues $\lambda=\pm 1$ (both with $m=1$). 

\begin{figure}[h]
\begin{center}
\includegraphics[width=0.48\textwidth]{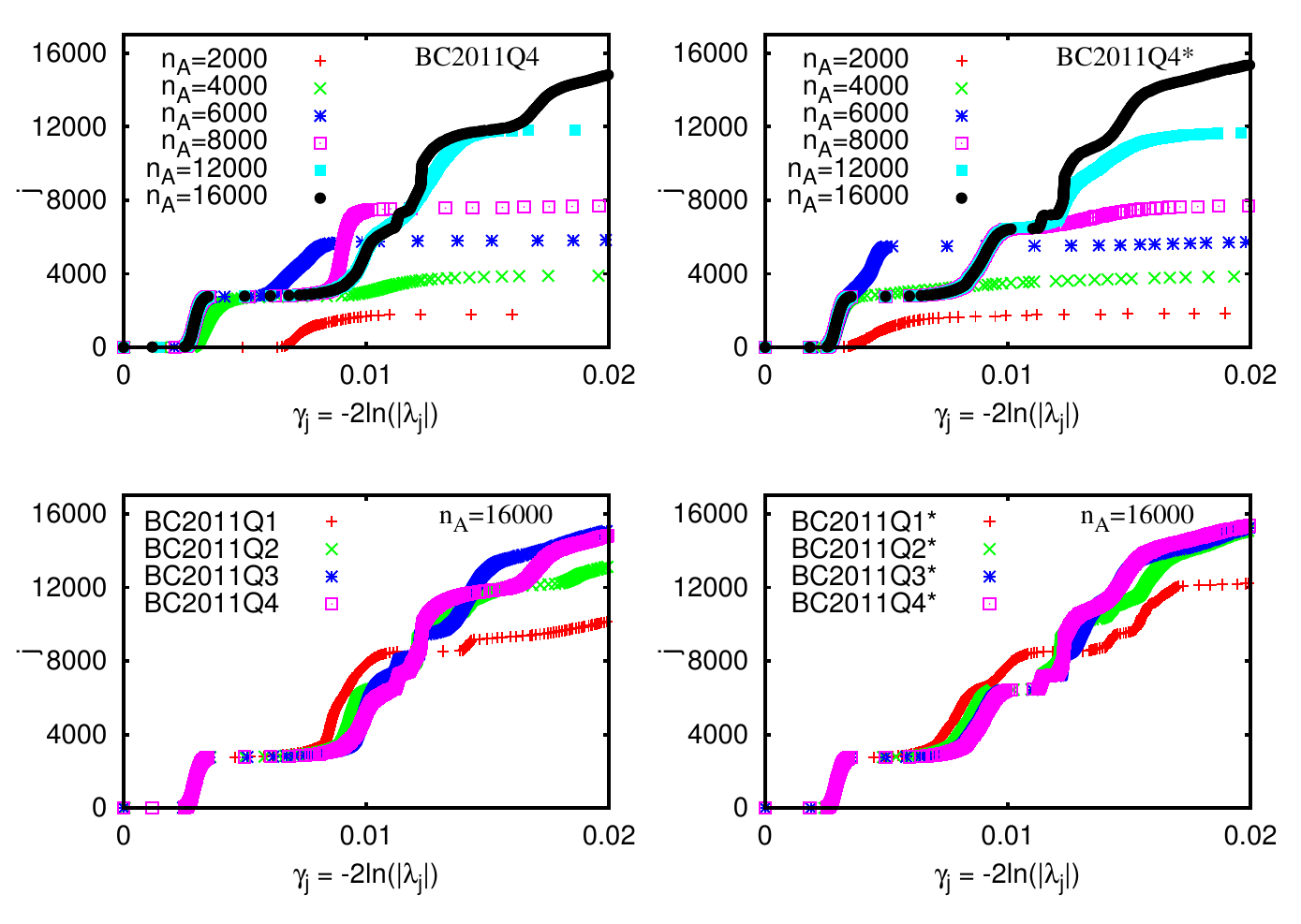}
\caption{(Color online)
Level number $j$ versus the decay width 
$\gamma_j=-2\ln(|\lambda_j|)$ with $\lambda_j$ being the 
$j$-th core space eigenvalue computed by the normal precision 
Arnoldi method with uniform initial vector and Arnoldi dimension $n_A$.
The top left (right) panel corresponds to the network BC2011Q4 
(or BC2011Q4*) for different values of $n_A$ with 
$2000\le n_A\le 16000$. 
The bottom left (right) panel corresponds to the four networks BC2011Qk, 
(or BC2011Qk*) for $k=1,2,3,4$ and $n_A=16000$. 
}
\label{fig8}
\end{center}
\end{figure}

The convergence with the increase of the Arnoldi dimension $n_A$
is illustrated in the top panels of Fig.~\ref{fig8} for
BC2011Q4 showing the dependence $j(\gamma_j)$
where $\;\;\;$ $\gamma_j=-2\ln|\lambda_j|$ with $\lambda_j$ being the core space
eigenvalue. For $S$ and $S^*$ the comparison between the two maximal values 
$n_A=16000$ and $n_A=12000$ indicates that about $j \approx 5000-6000$ 
eigenvalues 
up to $\gamma \approx 0.01$ are reliable. However, we remind that 
the comparison of different computations 
for $n_A=16000$ shows that the number of reliable 
eigenvalues is actually higher $\approx 12000$ corresponding 
to $\gamma \approx 0.015$. 
The circle structure well visible in Fig.~\ref{fig7}
is responsible for appearance of large steps in the dependence $j(\gamma)$
well seen in Fig.~\ref{fig8}. A similar dependence $\gamma_j$ 
is also present for other quarters BC2011Q1, BC2011Q2, BC2011Q3 shown in bottom panels 
of Fig.~\ref{fig8}.

\begin{figure}[h]
\begin{center}
\includegraphics[width=0.48\textwidth]{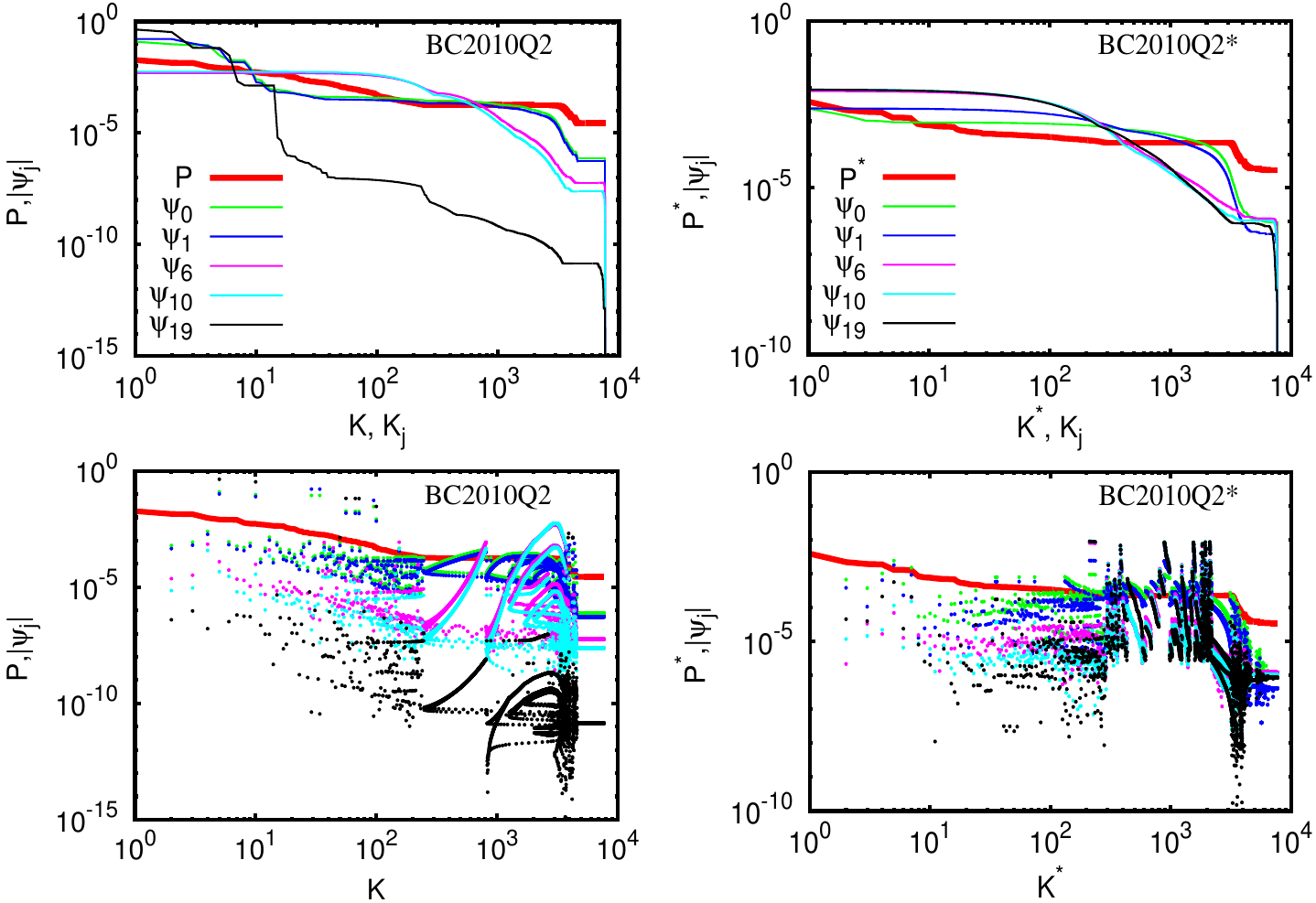}
\caption{(Color online)
{\em Top panels:} PageRank $P$ (or CheiRank $P^*$) at damping factor 
$\alpha=0.85$ and modulus $|\psi_j|$ of 5 selected eigenvectors of $S$ 
for the network BC2010Q2 (or BC2010Q2*) 
versus the index $K$ for PageRank $P$ ($K^*$ for CheiRank $P^*$) 
or the individual ordering index $K_j$ for each eigenvector $\psi_j$. 
{\em Bottom panels:} The same as top panels but only using the PageRank (CheiRank) 
index $K$ ($K^*$) on the $x$-axis for all shown vectors. 
Note that the given vector index values $0,1,6,10,19$ do 
not correspond to the level number of Fig.~\ref{fig8} but they 
correspond to an index of a selected set of 20 core space eigenvalues 
closest to the unit circle and with uniformly distributed eigenphases 
between 0 and $\pi$, i.e. the selected eigenvalues are roughly 
$\lambda_j\approx |\lambda_j|\exp(-i2\pi j/19)$ for $j=0,\dots,19$ 
and with $|\lambda_j|\approx 1$. In particular: 
$\lambda_0=0.99990030$ ($0.99999816$), 
$\lambda_1=0.99967849$ ($0.99612525+i 0.00704040$),
$\lambda_6=0.63708435+i 0.75027164$ ($0.63887779+i 0.75813987$), 
$\lambda_{10}=0.00130422+i 0.98315766$ ($-0.00043259+i 0.99066741$), 
and $\lambda_{19}=-0.99053491$ ($-0.99014386$) 
for BC2010Q2 (or BC2010Q2*). }
\label{fig9}
\end{center}
\end{figure}

\section{Eigenstates of BCN Google matrix}

The decay of PageRank and CheiRank probabilities 
at different quarters is presented in Fig.~\ref{fig4}.
Here we describe the properties of several eigenstates.
As soon as the eigenvalues are determined the eigenstates
corresponding to the selected eigenvalues can be efficiently
computed numerically as described in \cite{rmp2015,ulamfrahm,wikispectrum}.

\begin{figure}[h]
\begin{center}
\includegraphics[width=0.48\textwidth]{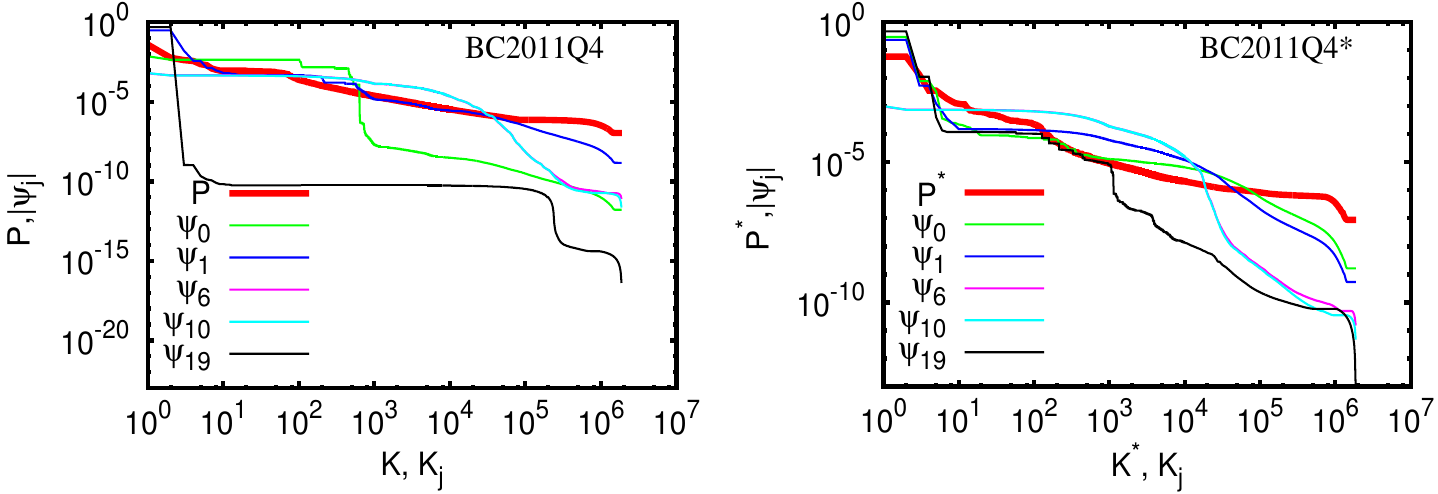}
\caption{(Color online)
PageRank $P$ (or CheiRank $P^*$) at damping factor 
$\alpha=0.85$ and modulus $|\psi_j|$ of 5 selected eigenvectors of $S$ 
for the network BC2011Q4 (or BC2011Q4*) 
versus the index $K$ for PageRank $P$ ($K^*$ for CheiRank $P^*$) 
or the individual ordering index $K_j$ for each eigenvector $\psi_j$. 
The given vector index is similar to Fig.~\ref{fig9}. 
The eigenvalues of the shown eigenvectors are: 
$\lambda_0=0.99999999$ ($0.99999996$),
$\lambda_1=0.99999322$ ($0.99907654$),
$\lambda_6=0.64270357+ i0.76431760$ ($0.64271106+ i0.76429326$), 
$\lambda_{10}=-0.00116375+ i0.99865094$ ($-0.00117162+ i0.99862722$),
and $\lambda_{19}=-0.99999317$ ($-0.99934321$) 
for BC2011Q4 (or BC2010Q4*). }
\label{fig10}
\end{center}
\end{figure}

The results for 6 eigenvectors of BC2010Q2 are shown in Fig.~\ref{fig9}
and for BC2011Q4 in Fig.~\ref{fig10}. The selected eigenvectors 
$\psi_0$, $\psi_1$, $\psi_6$, $\psi_{10}$, $\psi_{19}$ (additional to PageRank 
and CheiRank vectors) are marked by an index $j$ corresponding to 20 core eigenvalues
closest to the unit circle with uniformly distributed eigenphases between $0$ and $\pi$.
In the top panels of Fig.~\ref{fig9} we order all amplitudes $|\psi_j|$
in monotonically descending order with their own local-Rank index $K_j$
with maximum at $K_j=1$ ($K_j$ is different from PageRank index $K$).
The interesting feature ofs these eigenstates is the presence of 
large plateaus where for hundreds of nodes the amplitude $|\psi_j|$
remains practically independent of $K_j$. This indicates a presence of
relatively large communities of users coupled by certain links.
The bottom panels of Fig.~\ref{fig9} show the amplitudes
$|\psi_j|$ as a function of the global PageRank index $K$.
For the BC2010Q2 network the nodes with largest amplitudes  $|\psi_j|$
are located at relatively large $K$ values with $K<100$.
It is possible that these nodes correspond to bitcoin miners.
However, a significant number of nodes with
relatively large amplitudes are  located
at very high values $K \sim 2000$. For the Google matrix $G^*$
all large amplitudes are located at large values of the CheiRank index
$K^* > 500$. For the larger BC2011Q4 network, shown in Fig.~\ref{fig10}
we find the presence of similar plateau structure for eigenstate amplitudes.

\begin{figure}[h]
\begin{center}
\includegraphics[width=0.48\textwidth]{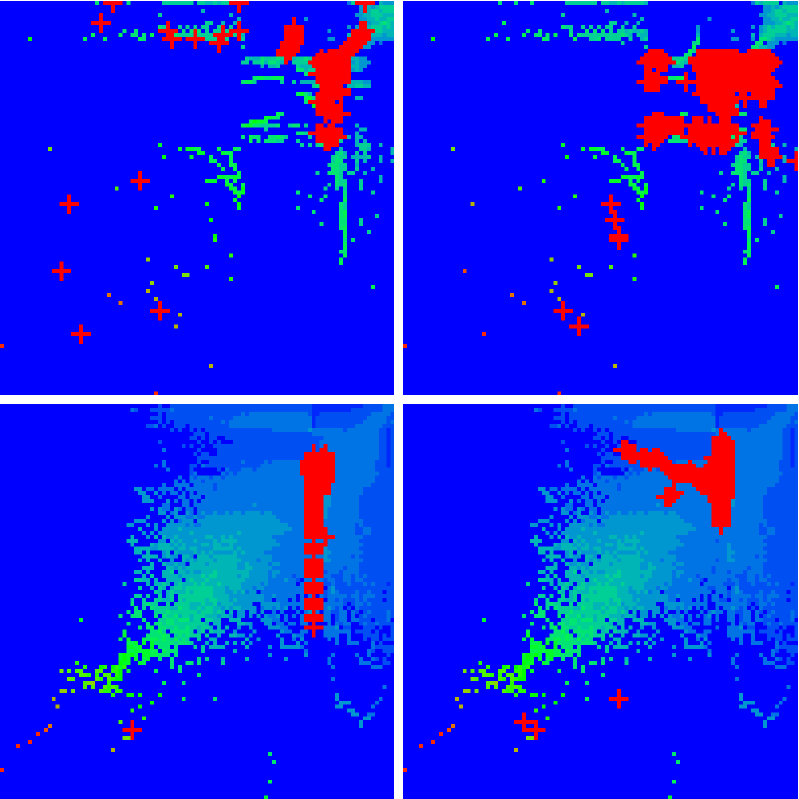}
\caption{(Color online)
Density of nodes  $W(K,K^*)$ on PageRank-CheiRank plane $(K,K^*)$
averaged over $100\times100$ 
logarithmically equidistant grids for $0 \leq \ln K, \ln K^* \leq \ln N$,
the density is averaged over all nodes inside each cell of the grid,
the normalization condition is $\sum_{K,K^*}W(K,K^*)=1$.
Color varies from blue at zero value to red at maximal density value.
In order to increase the visibility large density values have 
been reduced to (saturated at) 1/16 of the actual maximum density.
At each panel the $x$-axis corresponds to $\ln K$ 
and the $y$-axis to $\ln K^*$. 
Both top panels correspond to BC2010Q2 for K and to BC2010Q2* for $K^*$ 
and both bottom panels to BC2011Q4 for K and to BC2011Q4* for $K^*$. 
The red crosses show the top 1000 nodes of the eigenvector $\psi_{10}$ 
used in Figs.~\ref{fig9} and \ref{fig10} of 
BC2010Q2 (top left), BC2010Q2* (top right), BC2011Q4 (bottom left) and 
BC2011Q4* (bottom right). 
}
\label{fig11}
\end{center}
\end{figure}

\begin{figure}[h]
\begin{center}
\includegraphics[width=0.48\textwidth]{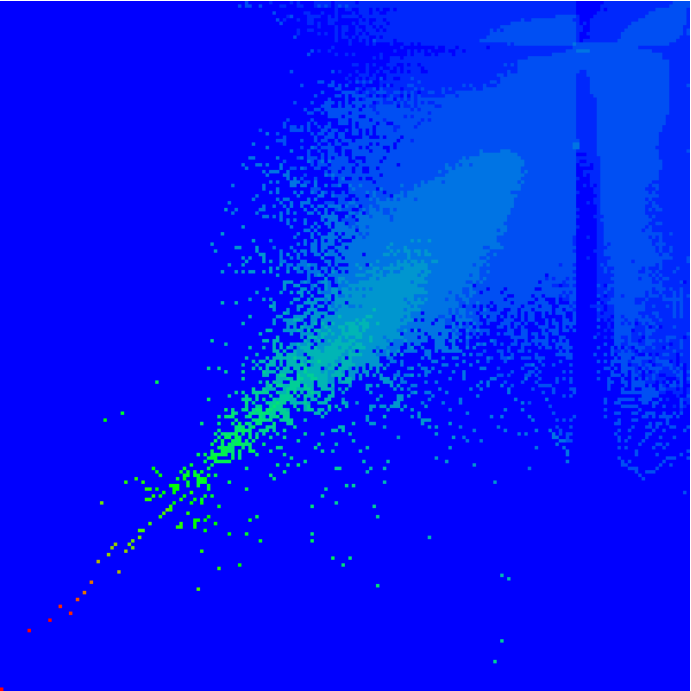}
\caption{(Color online)
Density of nodes  $W(K,K^*)$ on PageRank-CheiRank plane $(K,K^*)$ 
for BC2013Q2 averaged over $200\times 200$ 
logarithmically equidistant grids for $0 \leq \ln K, \ln K^* \leq \ln N$,
the density is averaged over all nodes inside each cell of the grid,
the normalization condition is $\sum_{K,K^*}W(K,K^*)=1$.
Color varies from blue at zero value to red at maximal density value.
In order to increase the visibility large density values have 
been reduced to (saturated at) 1/16 of the actual maximum density.
At each panel the $x$-axis corresponds to $\ln K$ 
and the $y$-axis to $\ln K^*$. 
}
\label{fig12}
\end{center}
\end{figure}

Similarly to Wikipedia and other networks \cite{rmp2015}
it is convenient to present the distribution
of network nodes on the CheiRank-PageRank plane $(K,K^*)$
shown in Fig.~\ref{fig11} for the cases of BC2010Q2 of 
Fig.~\ref{fig9} and BC2011Q4 of Fig.~\ref{fig10}. 
We see that for BC2010Q2 the density distribution of $N$ nodes
on $(K,K^*)$-plane is still strongly fluctuating,
but for BC2011Q4 it starts to stabilize and 
becomes close to the density of our largest network of BC2013Q2
shown in Fig.~\ref{fig12}. The important feature of the 
stabilized density distributions of BC2011Q4 and  BC2013Q2
is the fact that the maximum of distribution is located at the diagonal
$K=K^*$. This is similar to the situation of
the world trade network \cite{wtn1,wtn2} where each country (node)
or user for BCN tries to keep trade balance between
outgoing (export) and ingoing (import)
flows. 

In Fig.~\ref{fig11} we show by red crosses the location
of  top largest amplitudes $|\psi_j|$ at $j=10$ 
for Google matrices $G$ (left column) and
$G^*$ (right column). We see that only a few large amplitudes
are located at leading (smallest) values of $K$ and $K^*$.
This shows that the vector $\psi_{10}$ corresponds to a certain
rather isolated community.

\begin{figure}[h]
\begin{center}
\includegraphics[width=0.48\textwidth]{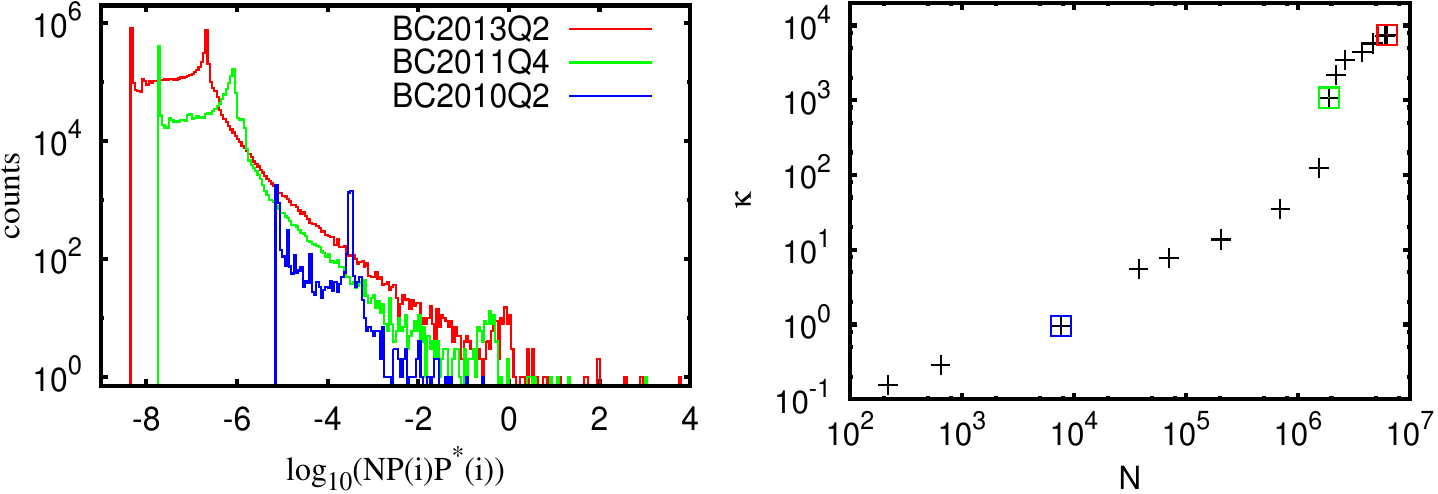}
\caption{(Color online)
{\em Left panel: } 
Histogram of frequency appearance of   
correlator components $\kappa_i=N P(K(i)) P^*(K^*(i))$ 
for the three networks BC2013Q2 (red), BC2011Q4 (green) and 
BC2010Q2 (blue).
For the histogram 
the whole interval  $10^{-{9}} \leq \kappa_i \leq 10^4$ is divided in 
260 cells of equal size in logarithmic scale.
{\em Right panel:} The dependence of the 
correlator $\kappa =N \sum^N_{i=1} P(K(i)) P^*(K^*(i)) - 1$
on the network size $N$ for all bitcoin networks between 
BC2009Q4 and BC2013Q2. The three data points surrounded by a colored square 
box correspond to the three networks of the left panel with the same colors.}
\label{fig13}
\end{center}
\end{figure}

The proximity of the density distribution
to the diagonal $K=K^*$ leads to a significant correlation between
PageRank and CheiRank vectors $P(K(i))$ and $P^*(K^*(i))$.
This correlation is convenient to characterized by 
the correlator \cite{rmp2015,linux,wikispectrum}
$\kappa = N \sum_{i=1}^{N} P(K(i)) P^*(K^*(i)) - 1$.
The large values of $\kappa$ corresponds
to a strong correlation of PageRank and CheiRank probabilities,
while $\kappa$ close to zero or even slightly negative 
appears to uncorrelated vectors $P$ and $P^*$.
The dependence of $\kappa$ on the network size 
$N$ is shown in Fig.~\ref{fig13} (right panel) where 
the correlator is becoming very large up to 
$\kappa \approx 10^4$ for the last quarters of BCN.
The frequency distribution of correlator components 
$\kappa_i=N P(K(i)) P^*(K^*(i))$ 
for three cases at different quarters is shown in the left panel 
of Fig.~\ref{fig13}.
These distribution show the presence of very active users with large
$\kappa_i$ values corresponding to their expected high 
activity of bitcoin outgoing and ingoing transactions.
It may be rather interesting to determine the hidden identity
of users with largest $\kappa_i$ values.

\section{Gini coefficient of BCN}

In economy the distribution of wealth of a certain population is
often characterized by the Gini coefficient proposed in 1912 
(see e.g. \cite{gini1912,wikigini,yakovenko}).
The Gini coefficient is typically defined using the {\em Lorenz curve}
which plots the fraction $y$ of the total income of a 
fraction $x$ of the population with the lowest income versus $x$. 
The line at $y=x$ thus represents perfect equality of incomes. 
The Gini coefficient is the ratio of 
the area that lies between the line of equality and 
the Lorenz curve normalized by the total area under the line of equality. 
Therefore the Gini coefficient is 0 for perfect equality and 
1 for complete inequality.

We can generalize this definition to PageRank and CheiRank distributions. 
For this let $P(K)$ be the usual PageRank vector with $K=1$ for the maximum 
value corresponding to the top PageRank node. Then we define the 
inverted PageRank as 
$P_{\rm inv}(K^\prime)=P(N-K^\prime-1)$ such that for $P_{\rm inv}$ 
the maximum value 
corresponds to $K^\prime=N$. In this way 
$P_{\rm inv}(K^\prime )$ represents in a certain way 
the ``income'' and its argument $K^\prime$ corresponds 
to the network nodes ordered in increasing order by their income 
(with lowest ``income'' for $K^\prime=1$ and maximum ``income'' 
for $K^\prime=N$). Then the cumulative income 
up to node $K^\prime$ is given by~:
\begin{eqnarray}
\label{gini1}
\sigma(P)_{K^\prime}&=&\sum_{\tilde K=1}^{K^\prime} P_{\rm inv}(\tilde K)=
\sum_{\tilde K=N-K^\prime+1}^N P(\tilde K). 
\end{eqnarray}
The notation $\sigma(P)$ reminds that $\sigma$ is defined 
with respect to a given PageRank vector (or CheiRank vector $P^*$ 
by replacing $P\to P^*$ in (\ref{gini1})). 
The quantity $\sigma(x)=\sigma(P)_{K^\prime}$ with $x=K^\prime/N$ corresponds 
to the standard Lorenz curve \cite{wikigini,yakovenko}. 
Therefore the Gini coefficient, defined as the area between 
$\sigma(P)$ and the line of equality 
normalized by the area below the line of equality  \cite{wikigini}, 
is given by:
\begin{eqnarray}
\label{ginig}
g&=&1-2\sum_{K^\prime =1}^N \sigma(P)_{K^\prime}/N =1- 2\sum_{K=1}^N K P(K)/N.
\end{eqnarray}
The Gini coefficient for the CheiRank $P^*$ is obtained in a 
similar way by using $\sigma(P^*)$ and replacing $P\to P^*$ 
in (\ref{ginig}). 

\begin{figure}[h]
\begin{center}
\includegraphics[width=0.4\textwidth]{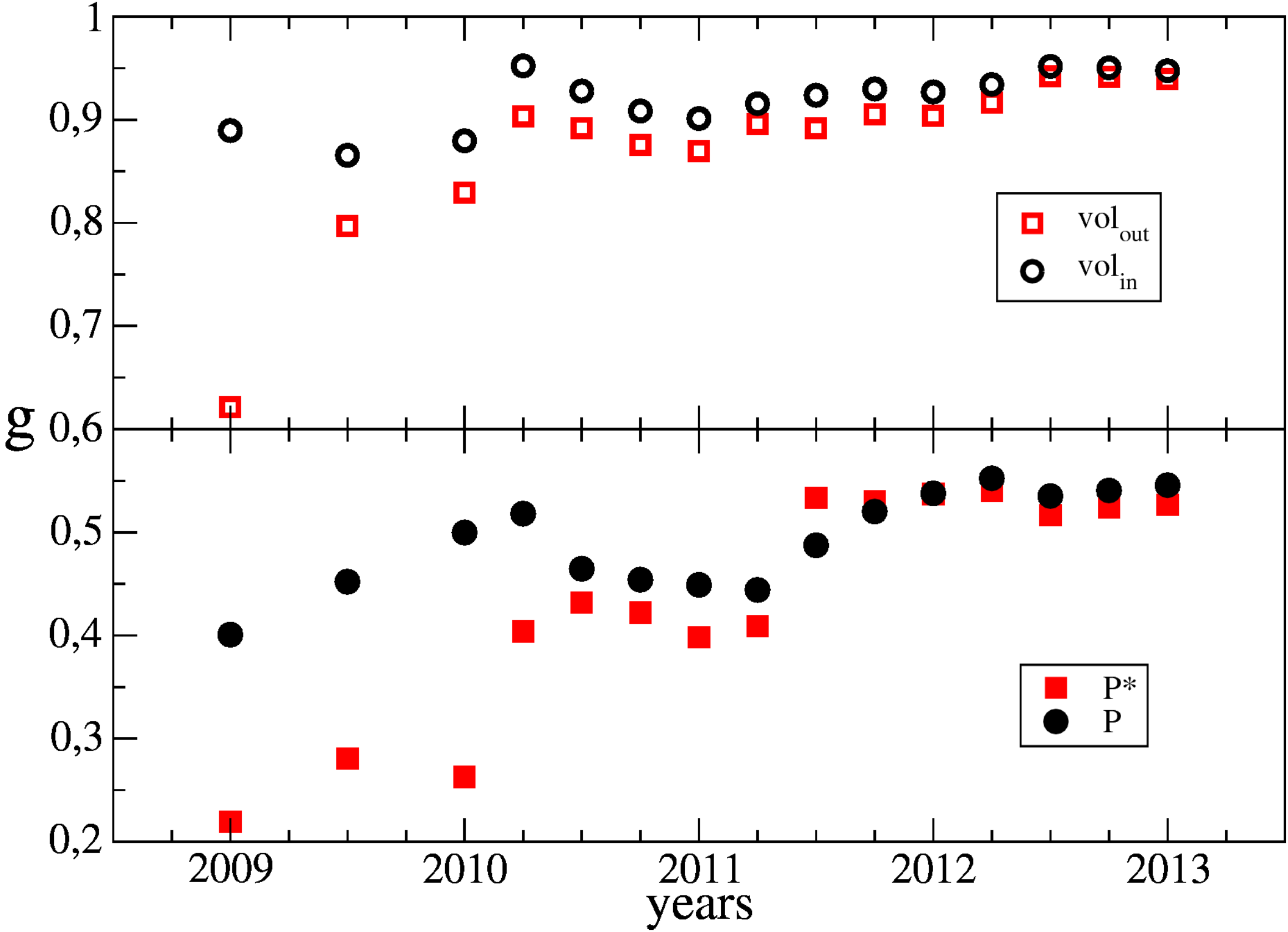}
\caption{(Color online)
{\em Top panel:} Gini coefficient $g$ time evolution for ingoing and outgoing bitcoin volume transfer
for quarter of years (halves for 2009).
{\em Bottom panel:} Gini coefficient $g$ time evolution for PageRank and CheiRank
of BCN for quarter of years (halves for 2009).
}
\label{fig14}
\end{center}
\end{figure}

The above definition of $g$ is done via the 
PageRank and CheiRank probabilities, i.e. where ``income'' corresponds 
to the PageRank or CheiRank values.
We will compare the corresponding $g$ values also with the standard definition
considering ingoing and outgoing amount of bitcoins (volume transfer) 
for BCN nodes (users) during a given quarter. 
The dependence of the different Gini coefficients,
defined via bitcoin volume transfer or PageRank and CheiRank probabilities,
on time is shown in Fig.~\ref{fig14}.

\begin{figure}[h]
\begin{center}
\includegraphics[width=0.4\textwidth]{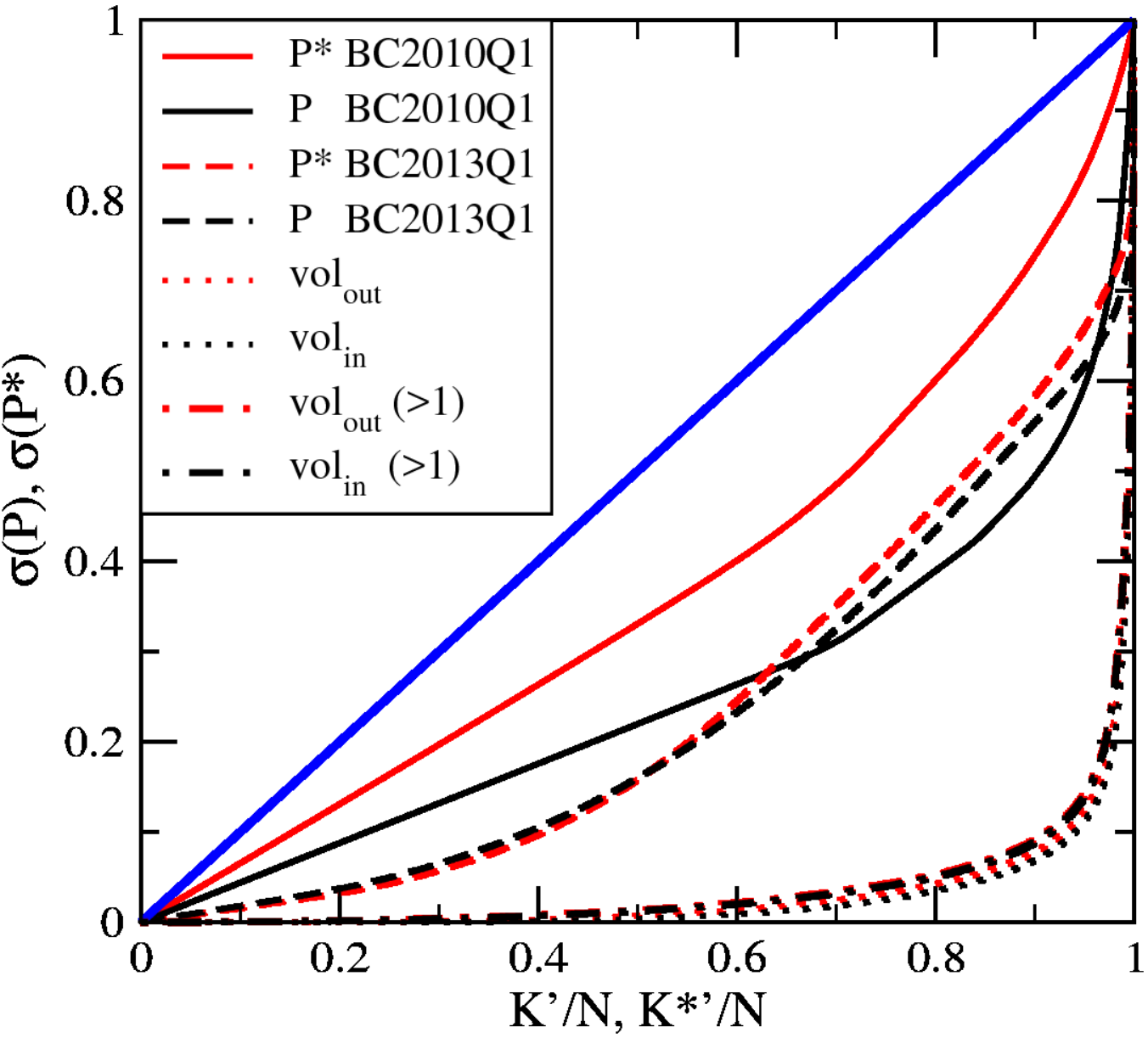}
\caption{(Color online)
The Lorenz curve of the bitcoin users (nodes) showing 
$\sigma(P)_{K^\prime}$ (and $\sigma(P^*)_{K^{*\prime}}$) vs. $K^\prime/N$ 
($K^{*\prime}/N$)
for PageRank and CheiRank of Q1 of 2010 and Q1 of 2013 (solid and dashed lines respectively).
The Lorenz curves for cumulative volume transfer are also shown  
for the full period 2009-2013 for out-going and in-going flow, taking into account users
with a minimum amount of 0 and 1 bitcoins (dotted and dot-dashed lines respectively).
The Gini coefficient values for volume transfer 
are $g=0.948$ (in-going larger than 0),  
$g=0.939$ (out-going larger than 0),
$g=0.927$ (in-going larger than 1),
$g=0.925$ (out-going larger than 1). 
The blue solid line represents the curve of perfect equality.
}
\label{fig15}
\end{center}
\end{figure}

For the BCN the evolution of Gini coefficient $g$, defined by (\ref{ginig}),
is shown in the bottom panel of Fig.~\ref{fig14}. 
We find that the Gini coefficient defined via volume (top panel 
of Fig.~\ref{fig14}) is stabilized from 2010
and takes a very high value $g \approx 0.9$. Such a large value of $g$ for 
bitcoin flows corresponds
to an enormously unequal wealth distribution between users \cite{wikigini,yakovenko}:
a small group of users controls almost all wealth.

We obtain smaller values of $g \approx 0.5$ for  PageRank and CheiRank probabilities.
We see that after 2010 the values of $g$ 
from PageRank and CheiRank probabilities become comparable. 
This corresponds to the 
stabilization of the node distribution in the PageRank - CheiRank plane
(see Fig.~\ref{fig11}, Fig.~\ref{fig12} discussed in the previous Section).
After 2010 we find $g \approx 0.5$ corresponding to a rather usual
value of $g$ with an exponential wealth distribution in a society
(see e.g. \cite{yakovenko}).
Also the Lorenz curve in 2013 (see Fig.~\ref{fig15})  becomes 
similar to USA income distribution (see e.g. Fig.8 in \cite{yakovenko}).

However, the above values are obtained with the PageRank and CheiRank probabilities
which are smoothing the row bitcoin flows due to the damping factor $\alpha$ in (\ref{eqGdefine}).
For the row bitcoin flows for the whole available period 2009-2013 we
find $g \approx 0.92$ (ingoing and outgoing $g$ values are rather close).
Such high $g$ values correspond to very unbalanced  wealth distribution
in the bitcoin community.

\section{Discussion}

We presented the results of Google matrix analysis
of bitcoin transaction network from the initial start in 2009 till April 2013.
From the period after 2010 the PageRank and CheiRank probability distributions 
are stabilized showing an approximate algebraic decay
with the exponent $\beta \approx 0.9$. We find that the spectrum of 
complex eigenvalues
of matrix $G$ has a very unusual form of circles being rather close
to the unitary circle. Such a structure has never appeared 
for other real networks
reported previously \cite{rmp2015}. The only example with a similar
spectral structure appears for the Ulam networks generated by intermittency maps
\cite{maps}. Such a circular structure corresponds to certain hidden communities coupled by
a long series of transactions. A manifestation of such communities with about hundreds of users
is also visible as a plateau structure in the eigenvectors of 
the Google matrix whose eigenvalues are close to the unit circle.
The distribution of users in the PageRank-CheiRank plain 
is maximal along the diagonal corresponding to the the fact that each user tries
to keep financial balance of his/her transactions.
A similar situation was also observed for the world trade networks \cite{wtn1,wtn2}.

We also characterized the wealth distribution for BCN users using the Gini coefficient $g$.
The definition of $g$ via PageRank and CheiRank probabilities leads to usual value $ g \approx 0.5$
for the time period after 2010 when the BCN is well stabilized.
However, the analysis of row bitcoin flows gives $g \approx 0.92$ 
corresponding to the situation when almost all wealth is
concentrated in hands of small group of users. We argue 
that the damping factor of the Google matrix
is responsible to a significant reduction of $g$ value.

Finally we note that the public access to all bitcoin transactions
makes this system rather attractive for analysis of statistical features
of financial flows. However, there is also a hidden problem of this network.
In fact it often happens that a user performing a transaction to another user
changes him/her bitcoin code after the transaction thus
effectively creating a new user even if the person behind the code remains the same.
This feature is responsible to the fact that the BCN is characterized by a rather low
ratio of number of links to number of nodes being about 2-3 while in
other networks like WWW and Wikipedia this ratio is about 10-20.
This low ratio value is at the origin of the strong sensitivity of eigenvalues of $G$
to numerical computational errors as we discussed in the paper.
Thus even if the bitcoin transactions are open to public it remains
rather difficult to establish the transactions between real persons.
In this sense the situation becomes similar to the transactions
between bank units: in this case the data are not public and are
hard to be accessed for scientific analysis. 

We note that all data used in our statistical analysis of BCN are available
at \cite{ourwebpage}.

We thank A.~D.~Chepelianskii and S.~Bayliss for stimulating discussions.
This research is supported in part by the MASTODONS-2017 CNRS project 
APLIGOOGLE (see \url{http://www.quantware.ups-tlse.fr/APLIGOOGLE/}).
This work was granted access to the HPC resources of 
CALMIP (Toulouse) under the allocation 2017-P0110. 


\end{document}